\documentclass[letterpaper, pre, twocolumn, floatfix]{revtex4}
\usepackage{amssymb,amsmath,bm,graphicx,subfigure}

\newcommand{\mb}[1]{\ensuremath{\mathbf{#1}}}
\newcommand{\figref}[1]{Fig.~\ref{#1}}

\newcommand{\eqnref}[1]{Eq.~\eqref{#1}}

\begin{document}
\title{Enslaved Phase-Separation Fronts in One-Dimensional Binary Mixtures}
\author{E.~M.~Foard and A.~J.~Wagner}
\affiliation{Department of Physics, North Dakota State University, Fargo, ND 58105}

\begin{abstract}
Phase-separation fronts leave in their wakes morphologies that are substantially different from the morphologies formed in homogeneous phase-separation.  In this paper we focus on fronts in binary mixtures that are \textit{enslaved phase-separation fronts}, \textit{i.e.} fronts that follow in the wake of a control-parameter front.  In the one-dimensional case, which is the focus of this paper, the formed morphology is deceptively simple: alternating domains of a regular size.  However, determining the size of these domains as a function of the front speed and other system parameters is a non-trivial problem.  We present an analytical solution for the case where no material is deposited ahead of the front and numerical solutions and scaling arguments for more general cases.  Through these enslaved phase-separation fronts large domains can be formed that are practically unattainable in homogeneous one-dimensional phase-separation.
\end{abstract}

\maketitle

\section{Introduction}
Phase-separation fronts are critically important for the formation of phase-separation morphologies in many practical applications.  Two industrial examples are the domain formation in binary alloys \cite{schaffer-2000} and immersion precipitation membranes \cite{cheng-1999, akthakul-2005, zhou-2006}.  A binary iron-carbon alloy which is slowly cooled from one side of the sample below its critical temperature may form a structure of alternating carbon-rich and carbon-poor bands \cite{jacot-1998, huang-2006}.  The temperature change moving through the alloy induces phase-separation where it crosses a critical value.  A very different example is the formation of immersion-precipitation membranes.  Here a polymer-solvent mixture is immersed in a non-solvent.  Phase-separation is induced as the solvent diffuses out of the mixture and is replaced by the non-solvent.  Phase-separation starts at the initial interface.  From there a phase-separation front advances into the mixture.  Both of these examples have in common that a space and time dependant parameter, \textit{i.e.} the temperature in the binary alloy case and the solvent concentration in the immersion precipitation case, induce phase-separation. We call such a quantity a control parameter. The advance of the temperature or the solvent concentration then constitutes a control parameter front which is responsible for inducing phase-separation.

Despite its critical roll in the formation of complex structures, only limited attention has been paid to this phenomenon.  There are a few notable exceptions. Furukawa \cite{furukawa-1990} examined the phase-separation in two dimensional binary mixtures following a sharp control parameter front and noted that the morphology formed was strongly determined by the speed of the front, and that morphology may change quite suddenly if the front speed is changed.  Hantz \cite{ISI:000235736200080} rediscovered the front speed dependence of morphology using a novel rotating control parameter front where the front speed depends on distance from the axis of rotation.  Dziarmaga \cite{dziarmaga-2001} suggested a model that is very similar to the one presented in the current work as a model for phase-separation in the early universe, and found that for slow fronts the density of domain walls is linear with front speed, \textit{i.e.} domain size is proportional to the inverse of front speed.

When a control parameter front is not progressing too quickly a phase-separation front follows directly behind and therefore moves at the same speed.  We call those fronts enslaved phase-separation fronts.  In the opposite case, the phase-separation front will not keep up with the control parameter front.  This then results in a free phase-separation front moving into an unstable medium.  This special case has received a significant amount of attention and a good review is given by Saarloos \cite{vansaarloos-2003-386}.

Enslaved phase-separation fronts give rise to a host of complicated phenomena, as witnessed by the large variety of morphologies formed.  For example, those formed by the previously mentioned immersion precipitation membranes \cite{cheng-1999}.  These morphologies are found to depend strongly on processing conditions which correspond to different shapes and speeds of the control parameter fronts.  This is why we initially focus on the simplest example of enslaved phase-separation fronts.  Here we consider only one-dimensional abrupt (\textit{i.e.} not spatially extended) control parameter fronts in binary mixtures. This already gives rise to unexpected and non-trivial behavior as explained below.  In particular it is noteworthy that large domains can be formed efficiently, even though such domain sizes are practically unobtainable through homogeneous phase-separation in one-dimensional systems.

In the next section we review the theory of spinodal decomposition, as it is relevant to the understanding of the dynamics of phase-separation at the front.  We then define our model which consists of a Cahn-Hilliard equation with an underlying $\phi^4$ free energy.  The control parameter front enters though the time and space dependence of the parameters in the free energy.  We expect the details of the phase-separation dynamics to depend strongly on the shape of the control parameter profile and its dynamics.  To simplify our analysis in this paper we focus on the simplest case consisting of a control parameter front which is a sharp step moving with a constant velocity.  To numerically explore the dynamics of this model we developed a lattice Boltzmann method (LBM) simulation. Generic results of these simulations elucidate the general behavior of an enslaved phase-separation front.  As the control parameter front advances into the mixed region, phase-separation ensues.  Excess material is then diffused both ahead of the front and through the forming domain into the next domain.  This continues until switching occurs and a new domain of the opposite kind is formed.  This is a regular process that repeats, resulting in regular domains after a few cycles.

To predict the morphology we only need to find the dependence of this one length-scale as a function of the nondimensional parameters of the equation of motion. Despite its apparent simplicity this is still a non-trivial problem.  We show that this can be done analytically in the case where the mobility ahead of the front is negligible and the front is moving very slowly.  The main ingredients in the analytical solution are solving for the time dependence of the concentration right behind the front and the condition for the switching to a new domain. Together, those allow us to determine the switching time and hence the domain size.

Using the results from this special case as a foundation, we examined more general enslaved fronts.  When we investigated the effect of non-zero mobility ahead of the front we observed that the domain-size has an unexpected non-monotonic dependence on the mobility. We were able to qualitatively explain this behavior as a competition of the diffusive deposition of material ahead of the front and the switching condition.

With satisfactory results for this model in one dimension, we conclude with a discussion of possible extensions such as using the front speed to imprint a domain pattern, and a roadmap of future research of enslaved front systems.

\section{Phase-Separation and Fronts}
As previously mentioned, phase-separation occurs after some physical control parameter, such as temperature, changes so that the system becomes thermodynamically unstable.  The system forms two or more distinct coexisting phases, and the property which distinguishes the phases is called the order parameter.  We are concerned with systems that transition from a one-phase state to a two-phase state as a control parameter is varied.  Despite the fact that most theoretical work on phase-separation has focused on the homogeneous case, most practical cases exhibit phase-separation that does not occur everywhere-at-once, but starts at one or more initial points and spreads through the system.  The boundary between phase separated and non phase separated material is called the phase-separation front.  Likewise, when the control parameter changes inhomogeneously we refer to the boundary between the single phase and two-phase regions as the control parameter front.  The correlation of the phase-separation front and the control parameter front depends on the exact nature of the dynamics of the control parameter.

Consider a binary mixture of ${\mathcal A}$-type and ${\mathcal B}$-type material described by the free energy:
\begin{equation}
	F=\int\!\!dx \left[ \frac{a}{2}\phi^2 + \frac{b}{4}\phi^4 + \frac{\kappa}{2}(\nabla\phi)^2 + c\phi \right] \;.
	\label{free_energy}
\end{equation}
This is a general fourth-order expansion of more specific free energy expressions.  The cubic term which would normally appear has been scaled away, without loss of generality, by choosing the critical concentration to be zero.  The $c$ parameter, which can usually be neglected, can matter here because it may take different values on either side of the control parameter front.  The dynamics of phase-separation is diffusive and is described by the Cahn-Hilliard equation
\begin{eqnarray}
	\partial_t\phi=\nabla(m\nabla\!\mu)\;,
	\label{cahn-hilliard}
\end{eqnarray}
where the chemical potential is derived from the free energy as
\begin{eqnarray}
	\mu=\frac{\partial F}{\partial \phi} = a\phi + b\phi^3 - \kappa\nabla^2\phi + c \;.
	\label{chemical_potential}
\end{eqnarray}
Here $a$ is the control parameter that determines the stability of the system; $a>0$ corresponds to the one-phase region and $a<0$ corresponds to the two-phase region.  The order parameter $\phi=\rho_{\mathcal A}-\rho_{\mathcal B}$ is the concentration of the two component materials.  For homogeneous parameters the linear $c$ term can be neglected because it does not change the equilibrium properties or the dynamics of homogeneous systems.  However, when $c=c(x)$ it will introduce chemical potential gradients between the material on either side of the control parameter front.  Of the other parameters, $m$ is the mobility, $\kappa$ determines the interfacial energy cost, and the amplitude of the non-linear term $b$ determines the equilibrium values of the order parameter:
\begin{equation}
	\phi_{eq} = \pm \sqrt{\frac{-a}{b}}\;.
	\label{phi_equilibrium}
\end{equation}

We now discuss different phase-separation mechanisms described by this model.

\subsection{Homogeneous Control Parameter Change}
If the control parameter does change rapidly everywhere at once---the well studied homogeneous quench---no control parameter front exists.  If the system is also homogeneous in composition a phase-separation front will not form, in which case phase-separation occurs via spinodal decomposition or nucleation.

Another interesting situation occurs when the control parameter is changed homogeneously but gradually.  In such a situation, cascades of sequential phase-separation events are observed, as shown in the work of Vollmer \textit{et al.} \cite{vollmer-2007,cates-2003-361}.  Such phenomena are probably also exhibited for extremely extended enslaved control parameter fronts but are outside the scope of the current paper.

In the case of a homogeneous quench, typically spinodal decomposition is observed.  It manifests itself in an exponential change of the order-parameter from initial fluctuations towards one of the two equilibrium values.  The growth rate depends on the fluctuation's wavelength, and the fastest growing wavelength will outgrow the others and quickly dominate the morphology.  This fastest growing wavelength is called the spinodal wavelength, denoted $\lambda_{sp}$.  Though it is not often discussed in the analysis of homogeneous quenches, we refer to the reciprocal of the fastest growth rate as the spinodal time, denoted $t_{sp}$.

To derive the spinodal wavelength and time one performs a linear expansion of the Cahn-Hilliard equation \eqref{cahn-hilliard} around the initial concentration $\phi_{in}$, then Fourier transforms this linearized Cahn-Hilliard equation into $k$-space which results in the ODE:
\begin{equation}
	\partial_t \tilde{\phi}(k) = -m\left[ \left( a+3b\phi_{in}^2 \right) k^2+\kappa k^4 \right] \tilde{\phi}(k) = R(k)\tilde{\phi}(k) \;.
\end{equation}
The solution of this equation is the exponential growth of fluctuations where the growth rate depends on the angular wavenumber $k$:
\begin{equation}
	\tilde{\phi}(k) = e^{R(k)t}\;.
\end{equation}
The angular wavenumber with the fastest growth rate corresponds to the spinodal wavelength:
\begin{equation}
	\lambda_{sp}=\frac{2\pi}{k_{sp}}=2\pi\sqrt{\frac{2\kappa}{-(a+3b\phi_{in}^2)}}\;.
	\label{spinodal_length}
\end{equation}
The fastest growth rate corresponds to the spinodal time:
\begin{equation}
	t_{sp}=\frac{1}{R(k_{sp})}=\frac{4\kappa}{m( a+3b\phi_{in}^2)^2}\;.
	\label{spinodal_time}
\end{equation}
Such a morphology, with an initial domain wavelength of $\lambda_{sp}$, will coarsen to larger length-scales, but the process is inefficient in one dimension since it obeys a logarithmic growth law $L\propto\ln(t)$ at late times \cite{tatsuzo-1986}.  Therefore it is  hard to form large structures in a finite time.

From the spinodal wavelength and time we define a spinodal speed
\begin{equation}
	u_{sp}=\frac{\lambda_{sp}}{t_{sp}}=\frac{\pi
          m}{\sqrt{2\kappa}}\left( -a-3b\phi_{in}^2 \right)^{3/2}\;,
	\label{spinodal_speed}
\end{equation}
which can be thought of as a natural speed of phase-separation.  This is important to our later analysis of the dynamics of enslaved phase-separation fronts because it allows us to define a nondimensional front speed.

\subsection{Free Propagation of a Phase-Separation Front}
Phase-separation fronts can form if a system has become unstable due to a sudden homogeneous control parameter change and a front is nucleated. This can occur if a defect causes local phase-separation to occur much more rapidly than through spinodal decomposition.  From this defect a front spreads through the system. Sarloos \textit{et al.} performed an analysis of the speed of such fronts for many different types of systems.  For a conserved order parameter system described by the Cahn-Hilliard equation \eqref{cahn-hilliard} with $a=-1$ and $m=b=\kappa=1$, the speed of the front, which they call the linear spreading speed, is:
\begin{equation}
	u^* = \sqrt{\frac{34+14\sqrt{7}}{27}} \left(1-3\phi_{in}^2\right)^{3/2} \approx 0.73 \; u_{sp}\;.
	\label{speading-speed}
\end{equation}
The initial domain wavelength is found to be smaller than the spinodal wavelength $\lambda^*\approx 0.35 \; \lambda_{sp}$ and the morphology then coarsens in time \cite{vansaarloos-2003-386}.

\subsection{Enslaved Phase-Separation Fronts}
Freely propagating phase-separation fronts can also form as a special case in a system with a control parameter front.  If the control parameter front moves at a speed $u\ge u^*$, this can result in the suppression of spontaneous spinodal decomposition, yet not inhibit the advancement of a freely propagating phase-separation front. However, if the control-parameter is moving slower than the linear spreading speed $u<u^*$, the phase-separation front cannot propagate freely and becomes enslaved by the control parameter front.

The nature of a control parameter front highly depends on the particulars of the physical system.  For instance, if temperature is the control parameter of a long binary metal alloy rod which is being cooled from one end, the control parameter front will resemble the familiar error-function solution of the heat-diffusion equation.  However, if the same rod is being extruded from a hot oven into a cooling environment, the control parameter front will be a much sharper transition. The shape and speed of the control parameter front will have an impact on the morphology formed in the wake of the front. In this paper we focus on phase-separation fronts enslaved by an abrupt control parameter front moving with a constant velocity.  This abrupt control parameter front will be represented by a step function for the parameters in the free energy and the mobility in the Cahn Hilliard equation.

We now introduce a lattice Boltzmann method to numerically investigate this system.  We will see that this control-parameter front induces phase-separation dynamics which is confined to a limited region around the front.  Consequently we can reduce the computational cost by focusing on a narrow region around the front.

\section{Lattice Boltzmann Method for Phase-Separation Fronts in 1D}\label{sec:LBM}
As we mentioned above, the dynamically important region for this model is restricted to a narrow area around the front.  So rather than simulate a stationary system with a front moving through it, we consider the equivalent case of material advected with velocity $u$ past a stationary phase-separation front.  This allows us to use a much smaller simulation size.  This changes the diffusive equation of motion \eqref{cahn-hilliard} to the advection-diffusion equation
\begin{equation}
	\partial_t\phi + \nabla(u\phi) = \nabla(m\nabla\!\mu)\;,
	\label{advectiondiffusion}
\end{equation}
where $u$ was the constant phase-separation front speed, which now appears as the material advection speed.  The parameters which define the abrupt control parameter front are
\begin{eqnarray}
	a(x)&=&a_S + (a_M - a_S) \Theta(x-x_0)\;,\\
	b(x)&=&b_S + (b_M - b_S) \Theta(x-x_0)\;,\\
	m(x)&=&m_S + (m_M - m_S) \Theta(x-x_0) \;, \\
	\kappa(x)&=&\kappa_S + (\kappa_M - \kappa_S) \Theta(x-x_0) \;,
	\label{parameter-set}
\end{eqnarray}
where $\Theta(x)$ is the Heaviside step function and $x_0$ is the position of the front.  Our naming convention uses the subscript $M$ to denote the mixing region ahead of the control parameter front, and the subscript $S$ to denote the separating region behind the control parameter front.

The LBM implementation of the advection-diffusion equation is similar to that of the ordinary diffusion equation, however the key differences of an advecting material with spatially dependent parameters warrents further clarification.

The standard lattice Boltzmann equation with the BGK approximation is
\begin{equation}
	f_i(x+v_i,t+1)-f_i(x,t)=\frac{1}{\tau}\left[f_i^0-f_i(x,t)\right]\;,
	\label{eqn:LBE}
\end{equation}
for discrete integer time, and a finite set of discrete velocities $v_i$ defined such that $x$ and $x+v_i$ are sites on the spatial lattice.  To determine the macroscopic evolution equation for this model we define a continuous $f_i\equiv f_i(x,t)$, and expand the first term to second order in a Taylor series as
\begin{equation}
	\nonumber f_i(x+v_i,t+1)=f_i + Df_i + \frac{1}{2}D^2 f_i+O(D^3)\;,
\end{equation}
where $D$ is the total derivative operator:
\begin{equation}
	D\equiv\partial_t+v_{i\alpha}\nabla_{\!\!\alpha}\;.
	\label{eqn:d}
\end{equation}
This allows us to rewrite \eqnref{eqn:LBE} as
\begin{equation}
	Df_i + \frac{1}{2}D^2 f_i +
        O(D^3)=\frac{1}{\tau}\left(f_i^0-f_i\right)\;,
	\label{eqn:LBE-almost_there}
\end{equation}
and to first order we obtain:
\begin{equation}
	f_i=f_i^0-\tau Df_i + O(D^2)\;.
	\label{eqn:chant-TO_FIRST_ORDER}
\end{equation}
With repeated use of \eqnref{eqn:chant-TO_FIRST_ORDER} to replace $f_i$ in \eqnref{eqn:LBE-almost_there} with the local equilibrium functions.  The hydrodynamic limit of the lattice Boltzmann equation then becomes
\begin{equation}
	Df_i^0 - D\left(\tau-\frac{1}{2}\right)Df_i^0 + O(D^3) =
        \frac{1}{\tau}\left(f_i^0-f_i\right)\;,
	\label{eqn:LBE-final}
\end{equation}
with the implicit generalization that that $\tau=\tau(x,t)$.  We then define the equilibrium moments
\begin{eqnarray}
	         \sum_i f_i = \sum_i f_i^0 &=& \phi \;, \\
	          \sum_i f_i^0 v_{i\alpha} &=& s u_\alpha\phi \;, \\
	\sum_i f_i^0 v_{i\alpha}v_{i\beta} &=& s \mu + s^2 u_\alpha u_\beta \phi \;,
\end{eqnarray}
corresponding to a conserved order parameter and a current of $su\phi$.  Summing up the Taylor expanded LBE \eqref{eqn:LBE-final} over $i$ we obtain the equation of motion
\begin{equation}
	s^{-1}\partial_t\phi + \nabla(u\phi) + O(\partial^3) = \nabla(m\nabla\!\mu)\;.
	\label{eqn:advectiondiffusion}
\end{equation}
This is \eqnref{advectiondiffusion} to second order, with the addition of a parameter $s$ which allows for rescaling simulation time.  The ability to scale simulation time allows us to trade computational speed for enhanced numerical stability and improved Galilean invariance \cite{wagner-2006}.  In this formulation the mobility $m$ is given by:
\begin{equation}
	m(x)=\tau(x) - \frac{1}{2}\;.
\end{equation}
In one dimension we use three discreet velocities
\begin{equation}
	v_i=\{0,-1,+1\}\;,
\end{equation}
which define the set of equilibrium distributions
\begin{eqnarray}
	f^0_0 &=& (1 - s^2 u^2)\phi - s \mu \;, \\
	f^0_{-1} &=& \frac{1}{2}\left((s^2 u^2 - s u)\phi  + s \mu \right) \;, \\
	f^0_{+1} &=& \frac{1}{2}\left((s^2 u^2 + s u)\phi  + s \mu \right) \;.
\end{eqnarray}
These equilibrium moments are then used with \eqnref{eqn:LBE} to calculate the time evolution of the system.

We now have to pay the cost for smaller simulation size allowed by fixing the position of the front in terms of more complicated inflow and outflow boundary conditions.  To represent a front moving in the positive $x$ direction we define $u<0$ which moves the material from right to left past the stationary front. This defines a fixed inflow boundary at $x=x_{max}$ and free outflow boundary at $x=0$.  Our inflow boundary condition is then given by
\begin{equation}
	f_{-1}(x_{max},t+1) = f_{+1}(x_{max},t) + s u \phi_{in} \;,
\end{equation}
which ensures a constant material influx $j_{in}=u\phi_{in}$.  On the other end of the simulation we have a free advection outflow boundary condition
\begin{equation}
	f_{+1}(0      ,t+1) = f_{-1}(0      ,t) - s u \phi(0,t) \;,
\end{equation}
ensuring an outflow of $j_{out}=u\phi(0)$.  Both boundary conditions are implemented as the densities are advected.  The boundary conditions are open and therefore the total integral of the order-parameter is not conserved.  To calculate the derivatives of the concentration required for the calculation of $\mu$ in \eqnref{chemical_potential} at the boundaries, we define the concentrations outside the lattice to be
\begin{eqnarray}
	\phi(x_{max}+1,t) &=& \phi_{in}     \;, \\
	\phi(-1,t) &=& \phi(0,t) \;.
\end{eqnarray}
This fully defines our lattice Boltzmann method.

\section{Dimensional Analysis}
Our main purpose is to predict the final morphology formed by enslaved phase-separation fronts.  In \figref{timelapse} we show the evolution of the phase-separation front and we see that only regular alternating domains are formed.  Thus, the final morphology is fully characterized by the length of the domains formed as a function of the parameters:
\begin{equation}
	\lambda =
        \lambda(a_M,a_S,b_M,b_S,m_M,m_S,\kappa_M,\kappa_S,c_M,c_S,u,\phi_{in})\;.
	\label{eqn:full-free-params}
\end{equation}
However, not all of these parameters are independent.  To reduce the number of free parameters we define the following nondimensional scales:
\begin{eqnarray}
		\label{eqn:nondimension_length}
	X&=&\frac{x}{\lambda_{sp}} = \frac{x}{2\pi}\sqrt{\frac{-a_S}{2\kappa_S}} \;,\\
		\label{eqn:nondimension_time}
	T&=&\frac{t}{t_{sp}}             = \frac{t m_S a_S^2}{4\kappa_S} \;, \\
		\label{eqn:nondimension_phi}
	\Phi&=&\frac{\phi}{\phi_{eq}}       = \phi \sqrt{\frac{b_S}{-a_S}} \;.
\end{eqnarray}
Here the length and time are scaled by their spinodal values, and the concentration is scaled by the positive equilibrium concentration of the separated material.  In contrast with homogeneous quench spinodal decomposition and free-front propagation, the initial nondimensional concentration ($\phi_{in}$) does not play a meaningful roll in the scaling of the dynamics of phase-separation for enslaved front systems.  We therefore set $\phi_{in}=0$ for the definition of spinodal wavelength in \eqnref{spinodal_length} and spinodal time in \eqnref{spinodal_time} used for nondimensionalization.

When we apply these nondimensionalizations to the equation of motion \eqref{advectiondiffusion} we obtain:
\begin{eqnarray}
	&& \partial_{T}\Phi
		+ \nabla_{\!\!X}\left( \frac{u}{u_{sp}} \Phi \right)
		= \frac{1}{2\pi^2} \nabla_{\!\!X} \left(
		\frac{m(x)}{m_S} \nabla_{\!\!X}
		\right. \\
	\nonumber && \left. \left(
		\frac{-a(x)}{a_S}\Phi
		+ \frac{b(x)}{b_S}\Phi^3
		- \frac{1}{8\pi^2}\frac{\kappa(x)}{\kappa_S} \nabla^2_{\!\!X} \Phi
		+ \frac{c(x)}{-a_S\phi_{eq}}
		\right) \right) \;.
\end{eqnarray}
This leaves us with the following dimensionless parameters:
\begin{eqnarray}
	\label{eqn:nondimension_speed} &&
		U = \frac{u}{u_{sp}} = \frac{u}{\pi m_S}\sqrt{\frac{2 \kappa_S}{(-a_S)^3}} \;,\\
	\nonumber &&
		M = \frac{m_M}{m_S} \;,\;
		A = -\frac{a_M}{a_S} \;,\;
		B = \frac{b_M}{b_S} \;,\\
	&&
		K = \frac{\kappa_M}{\kappa_S} \;,\;
		C = \frac{c_M-c_S}{a_S\phi_{eq}} \;.
	\label{nondimparams}
\end{eqnarray}

The main effect of $b_M$ is to renormalize $a_M$ and $c_M$, and the additional nonlinear effects can typically be neglected unless the concentration varies substantially in the mixing region.  We therefore consider $B$ an irrelevant nondimensional parameter and choose it to be zero.
We choose $C$ in such a way as to avoid bulk chemical potential differences between the mixed and separated regions $\mu^{bulk}_M(x\to \infty) = \mu^{bulk}_S(x\to -\infty)$, and we obtain 
\begin{equation}
	C = A\Phi_{in}\;.
\end{equation}
Additionally, we do not consider changes in the interfacial free energy cost across the front, and therefore set $K=1$.  This results in a nondimensional domain wavelength which is a function of four nondimensional parameters:
\begin{equation}
	L = L(A,M,U,\Phi_{in})\;.
	\label{eqn:freeparams}
\end{equation}

To investigate the dependence of $L$ on $U$ we first turn to numerical simulations.

\section{Simulation Results}
\begin{figure}
	\begin{center}
		\subfigure[Concentration $\phi(x,t)$]{\includegraphics[clip=true,width=\linewidth]{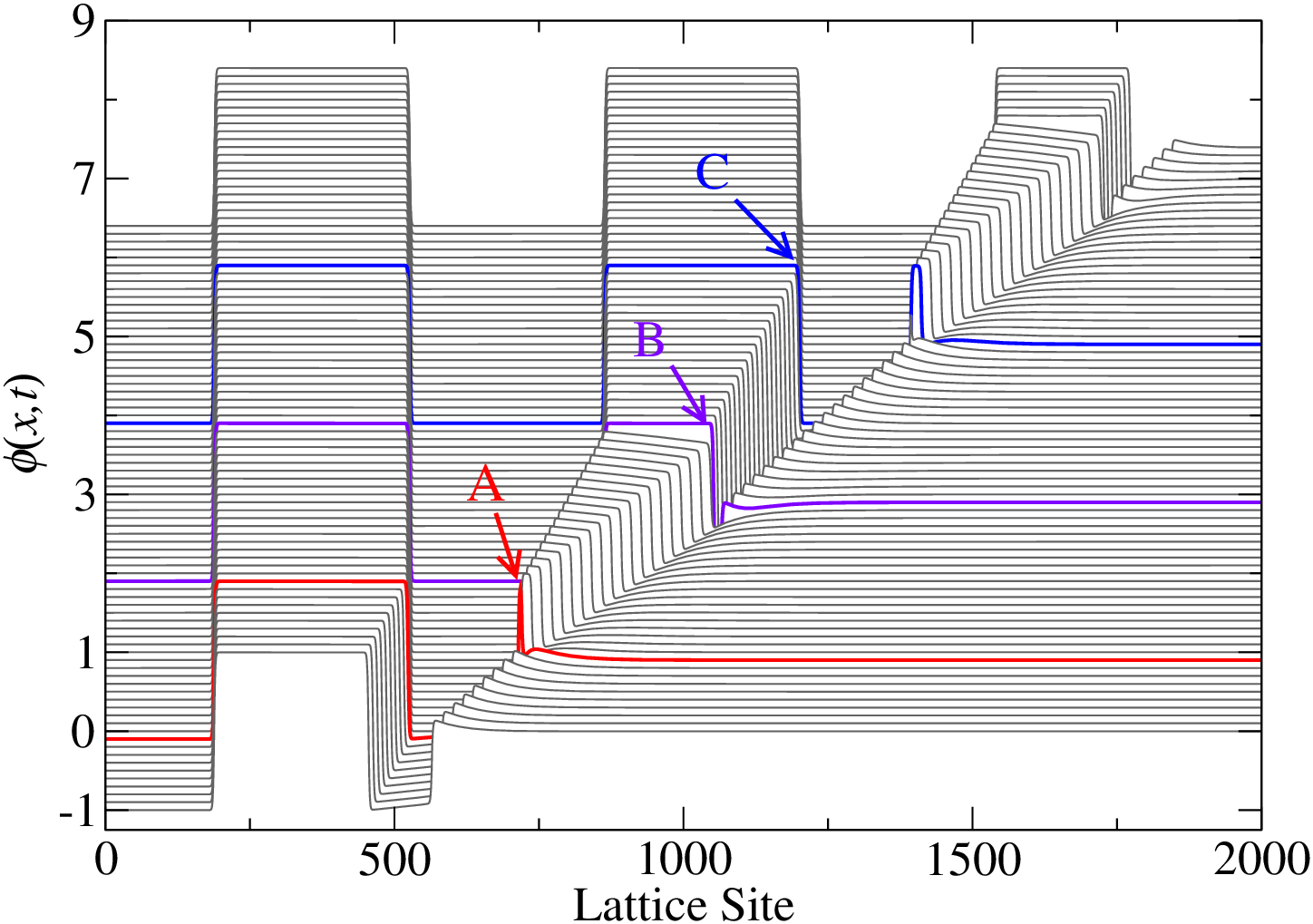}}
		\subfigure[Chemical Potential $\mu(x,t)$]{\includegraphics[clip=true,width=\linewidth]{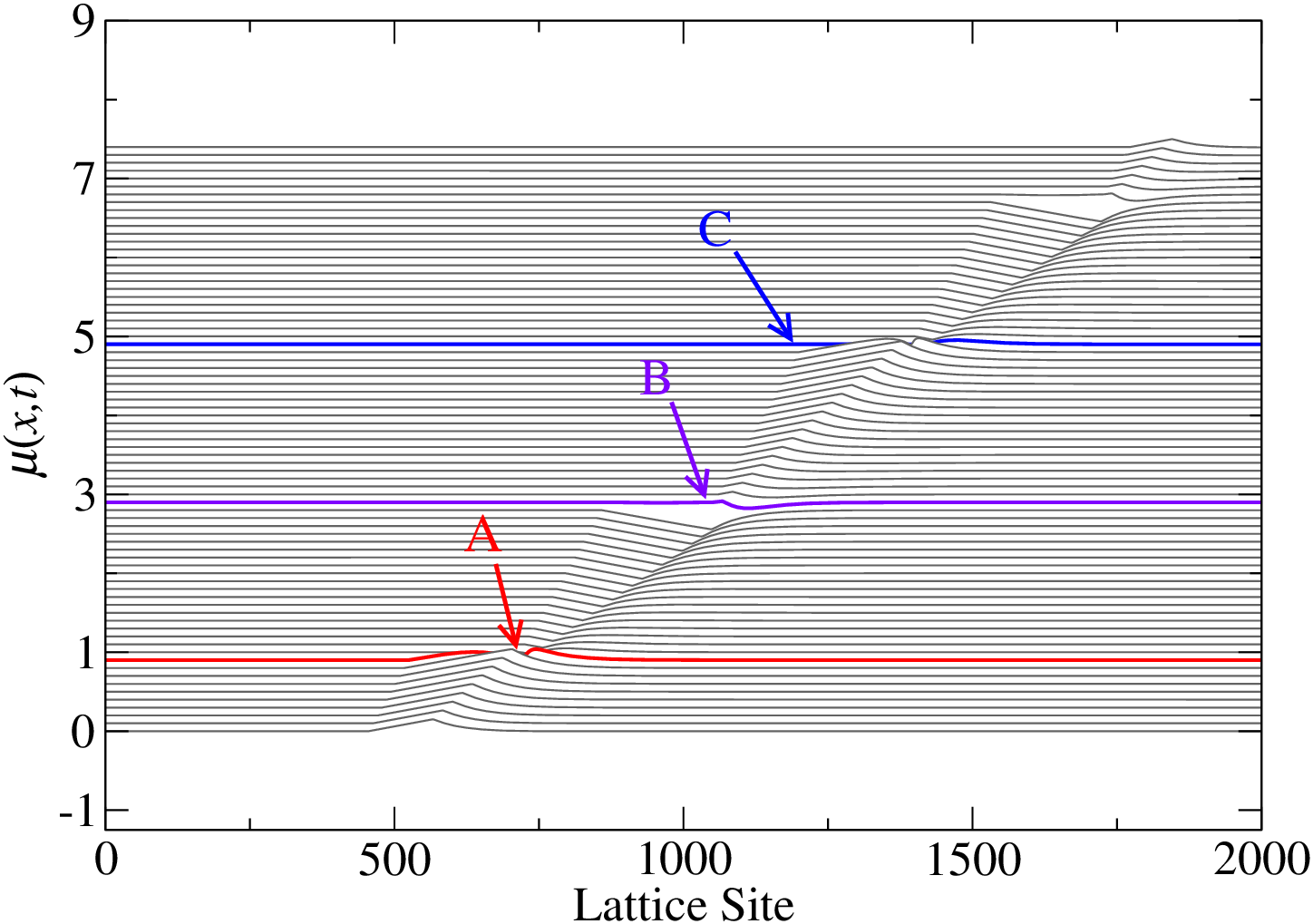}}
	\end{center}
	\caption{(color online)  Timelapse plots for an enslaved phase-separation front moving from right to left at nondimensional speed $U=0.001$ leaving regular alternating domains in its wake.  The first recorded profile is at the bottom of the graph with subsequent profiles shifted in the positive $y$-axis.  Each profile is separated by $200,\!000$ iterations, or a nondimensional time difference $\Delta T=12,\!500$. The $x$-axis scale is in lattice sites.  The upper graph shows the concentration profile over nearly two cycles of domain formation. The concentration ahead of the front can be seen to increase over time as the width of the two domains behind the front increases, until a new domain is nucleated.  The lower graph is the total chemical potential for this system, showing that the gradient of the chemical potential is flat except for just ahead of the front and across the first domain behind the front.  Additional simulation parameters are $A=1$, $M=0.1$, $\kappa=2$, $\phi_{in}=0$, and $s=0.1$.}
	\label{timelapse}
\end{figure}
Let us first consider a generic simulation to discuss the main features of morphology formation.  The result of one simulation using this one-dimensional LBM implementation is shown in \figref{timelapse} as a space-time plot of concentration and chemical potential.  It is clear from this plot that the dynamics of phase-separation is limited to the region around the front, and the chemical potential away from the front is flat.  This simulation started with equal parts ${\mathcal A}$-type and ${\mathcal B}$-type material in the mixed region, and the ${\mathcal A}$ and ${\mathcal B}$ domains formed after the front passed are therefore of equal size.

We shall now examine the domain formation using the middle ${\mathcal A}$-type domain in \figref{timelapse} as an example.  Three key events in the formation of this domain are marked A, B, and C\@.  At point A the domain is nucleated.  The concentration plot shows a very narrow domain and the chemical potential plot shows the previous peak collapsing toward a flat profile.  The domain there expands as it is pulled along with the front.  To grow this domain, ${\mathcal B}$-type material must be transported away.  From the gradient on the chemical potential we see that most of this material is pulled across and deposited on the back of the domain, which advances the back interface of the domain.  Some of the rest is pushed ahead of the front, and a small amount builds up on the forming domain just behind the front.  These last two depositions have the effect of making the forming domain increasingly unstable as the concentration at the front increasingly deviates from the equilibrium concentration.  To maintain the currents of material away from the interface, the chemical potential gradients must be maintained.  Therefore the trough in the chemical potential deepens.

At point B the ${\mathcal A}$-type domain has just detached from the front as a new ${\mathcal B}$-type domain has been nucleated.  The chemical potential very quickly becomes flat across the newly detached ${\mathcal A}$-type domain as it fully separates to the positive equilibrium concentration.  However, the formation of a ${\mathcal B}$-type domain at the front causes ${\mathcal A}$-type material to be deposited on the leading edge of the detached domain, thereby expanding it.  The interface between the ${\mathcal A}$-type domain and the new ${\mathcal B}$-type gets pulled along by the front, but is moving with about half the front's speed.  The rate of deposition of material on this moving interface is nearly constant as revealed by the almost constant linear slope of the chemical potential.

Point C marks the end of the formation of the domain by the enslaved phase-separation front as yet another domain is nucleated.  The chemical potential curve has become flat immediately adjacent to the domain, and all material currents to the interfaces have stopped.  The domain is now completely stationary.  Because the domains are highly ordered and much larger than the interface width, they do not evolve further at any appreciable rate \cite{tatsuzo-1986}.

We now qualitatively understand how the morphology is formed by a one-dimensional enslaved front.  To make quantitative statements we need to measure the relevant quantities from the LBM simulations of our model.

\subsection{Measurement Methods}
The sizes of domains formed are found numerically by determining the length between zero-crossings of the concentration.  To find the zero-crossing point to sub-lattice precision we linearly interpolate between the two lattice sites on either side of the crossing.  The first two domains after the control parameter front are still forming and expanding, therefore domain length measurements begin at the third zero-crossing behind the front and is an average of two or more domains.  To ensure that there are enough fully-formed domains, we dynamically grow the simulation to always keep the number of interfaces behind the front greater than those used to calculate a meaningful average.  The measurement is recorded once the uncertainty in the domain length is both less than one lattice spacing, and less than one thousandth the domain length itself.  To more directly compare our enslaved front system to previous work, we double the domain size to find the domain wavelength.  For asymmetric ($\phi_{in} \ne 0$) the domain size is multivalued and may lead to some confusion.  In this paper, however, we only present domain wavelength measurements for symmetricly mixed material.

In addition to the sizes of domains formed, we need to know the time-dependent concentration of the partially phase-separated material near the front.  We cannot simply measure the concentration at the front, as it would always be near zero due to the interface there.  However, since the chemical potential is continuous across the front we can invert the bulk chemical potential (defined as \eqnref{chemical_potential} for $\kappa=0$) to find the relevant concentration.  As shown in the timelapse plots of \figref{timelapse}, the chemical potential has a local extremum at the location of the control parameter front.  The chemical potential value we use to find the near-front concentration is the interlattice extremum value of a polynomial fitted to the three most-extreme values near the front.

\subsection{Domain Size as a Function of Front Speed}
\begin{figure}
	\subfigure[Domain sizes as a function of $u$ and $\kappa$.]{
		\label{univscaling_a}
		\includegraphics[clip=true,width=\linewidth]{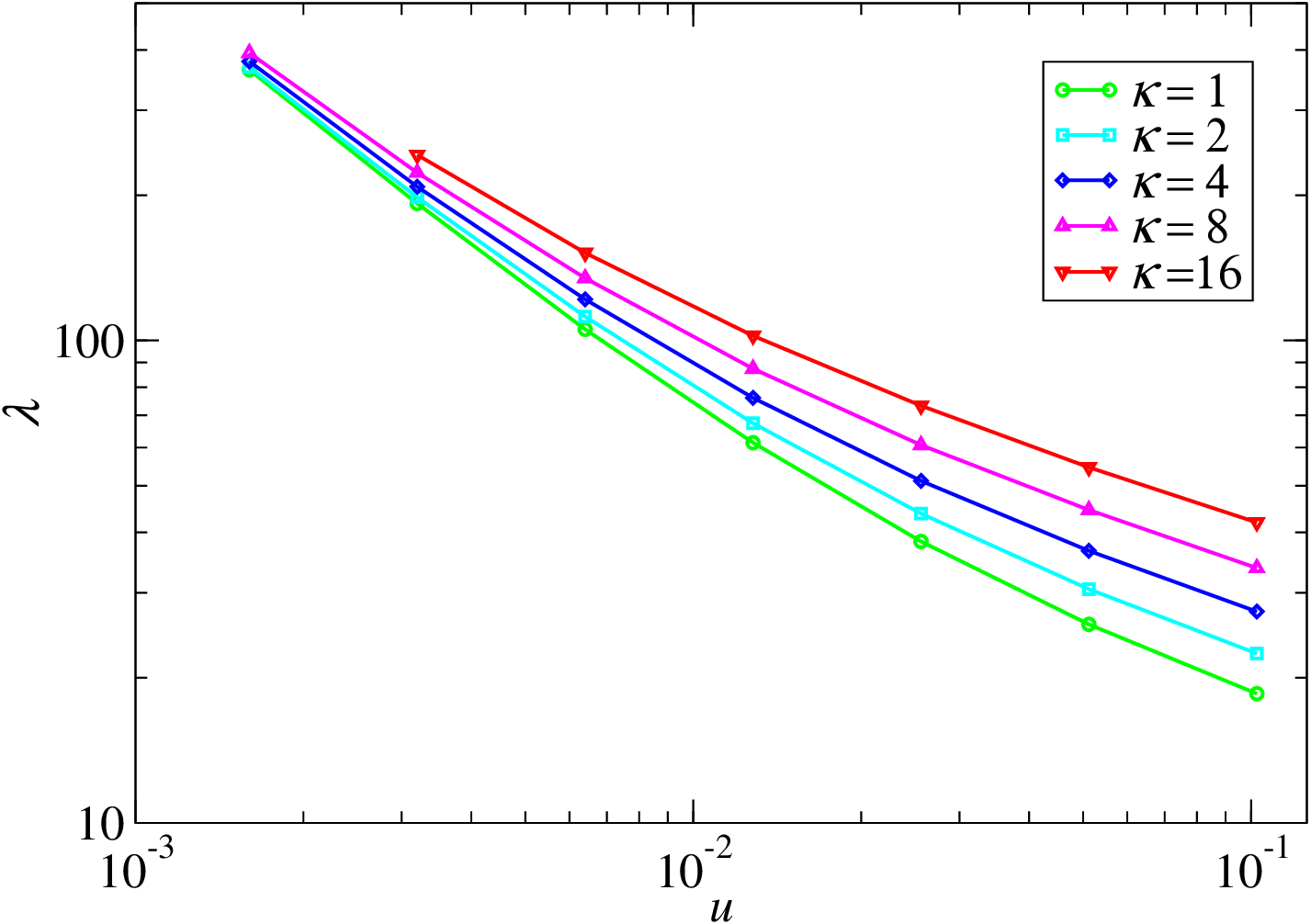}}
	\subfigure[Nondimensionalized domain wavelengths.]{
		\label{univscaling_b}
		\includegraphics[clip=true, width=\linewidth]{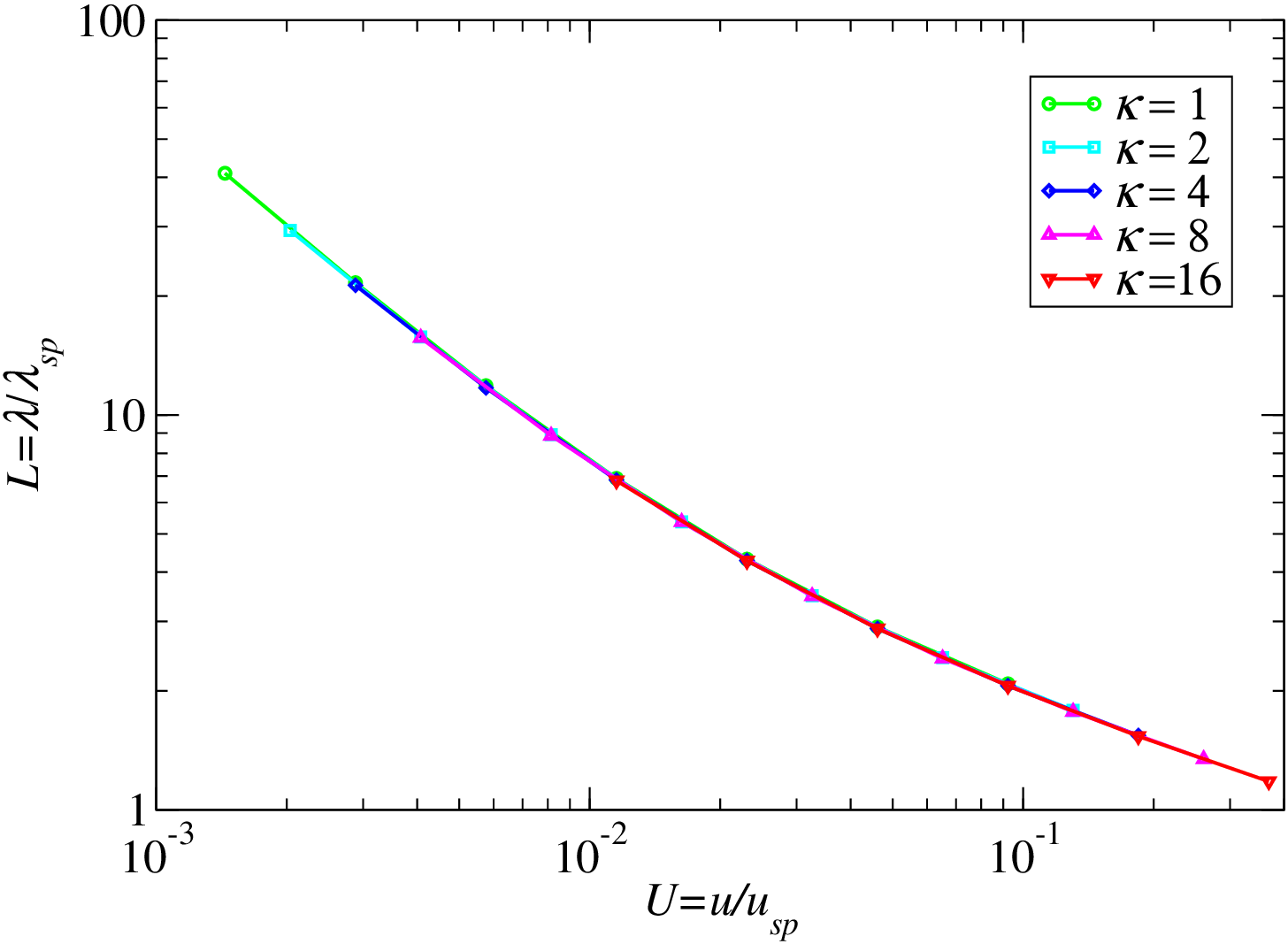}}
	\caption{(color online)  Simulation results for domain wavelength $\lambda$ as a function of phase-separation front speed $u$ for various interfacial energy costs.  Results are shown for raw data (top) and after rescaling to nondimensional lengths and speeds (bottom).  A single constant value of mobility $m_M=m_S=1/2$ was used for these simulations.}
	\label{univscaling}
\end{figure}
Since it is the speed of the control parameter front that determines whether the phase-separation front is enslaved, and for many physical systems the front speed may vary over time, it seems natural to first observe how the front speed determines the size of domains.  In addition, the interfacial free energy cost $\kappa$ plays an important roll in determining homogeneous and free-front length scales.  Therefore its effect on domain size will also be investigated.

As shown in \figref{univscaling_a}, domains get larger the slower the front becomes.  Also evident is that the larger the domains become the smaller the effect from the interfacial free energy, and that for slow enough front speeds the effect would become negligible.  However, we know from our nondimensionalization of the equation of motion that all dependence on the interfacial energy cost should scale away completely, but only if the simulation is capable of faithfully reproducing the equation of motion.  In \figref{univscaling_b} the nondimensional rescaling of the top plot is shown, revealing an impressive data collapse onto a single curve.

The data collapse provided by the nondimensionalization, though not unexpected, is very encouraging.  We can be confident that the appropriate parameters for continued investigation of this model are those revealed in \eqnref{nondimparams} by the nondimensionalization procedure.  In the remainder of this paper, we investigate the dependence of the domain wavelength on the nondimensional parameters given in \eqnref{eqn:freeparams}.  We now analytically investigate a situation where the nondimensional front speed $U$ is the only relevant nondimensional parameter.

\section{Analytical Solution for M=0}
To develop an analytical theory of domain formation we focus on the special case of $M=0$, where there is no dynamics in the mixed material ahead of the front.  This reduces the relevant parameters to just the enslaved front speed $U$.  All other dimensionless parameters are irrelevant since there is no dynamics ahead of the front.  Due to its simplicity, this system lends itself especially well to an analytical approach.  We will predict the resultant morphology (domain wavelengths) by explicitly solving the dynamics of phase-separation at the front.  We accomplish this in two parts: determine the time-dependence of the concentration just behind the front, and find the concentration at which a new domain will nucleate.  By combining these two results we find the switching time $t_{sw}$, and thus the domain wavelength:
\begin{equation}
	\lambda = 2 u t_{sw} \;.
	\label{eqn:domainlength}
\end{equation}

To numerically verify our analytical results we cannot simply set the mobility of the mixed region to zero.  This would require that $\tau_M=1/2$, which would lead to an unstable simulation \cite{chen-1998}.  Instead, the $M=0$ case is implemented by placing the control parameter front at the positive boundary, and defining the off-lattice concentration as $\phi_{in}$.

\subsection{Concentration at the Front}
\begin{center}
	\begin{figure}
		\includegraphics[clip=true, width=\linewidth]{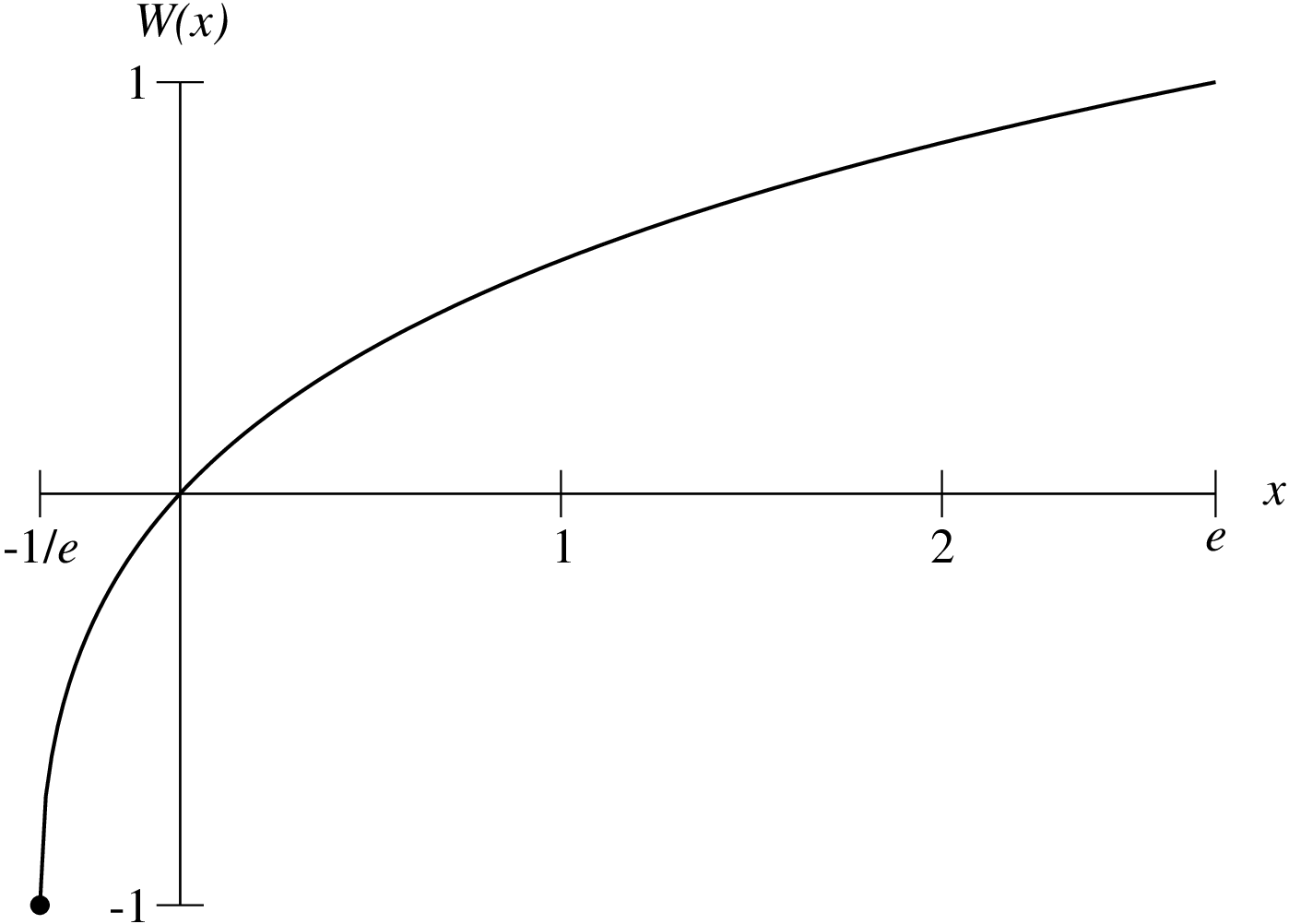}
		\caption{Real principle branch of the Lambert~W function for $1/e\leq x \leq e$ \cite{corless-1996}.  The Lambert~W function appears in the solution for an enslaved front with $M=\Phi_{in}=0$ for $U\ll 1$.}
		\label{fig:lambertW}
	\end{figure}
\end{center}
\begin{center}
	\begin{figure}
		\includegraphics[clip=true, width=\linewidth]{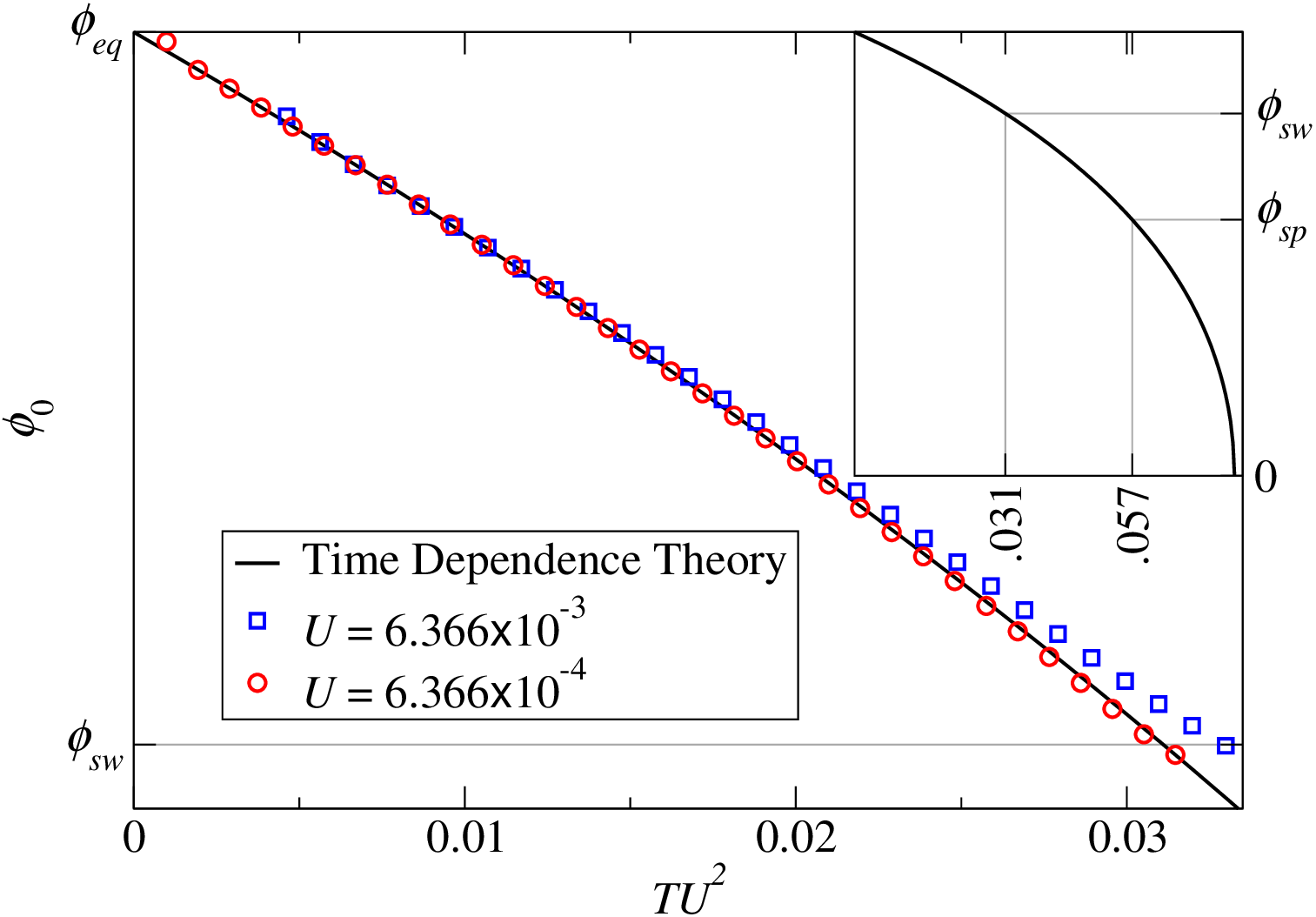}
		\caption{(color online)  Comparison of the simulation results for the time evolution of the concentration at the phase-separation front for fast and slow front speed with $\phi_{in}=0$.  Note that the fast phase-separation front speed simulation shows marked deviation from the theoretical curve, as expected.  Also note that the concentration at which the domain switches ($\phi_{sw}$) does not seem to depend on front speed.  Inset shows the full theoretical curve, though concentration should never go below $\phi_{sp}$ as that is unconditionally unstable and would immediately undergo spinodal decomposition.  While not essential to this analysis, zero concentration occurs for $T U^2 =(\ln(16)-2)/\pi^2\approx0.078$.}
		\label{fig:diffeq-phi-of-t}
	\end{figure}
\end{center}
To determine the time evolution of the concentration at the phase-separation front we examine the current of material into and out of the domain interface formed at the front.  We first assume the interface to be sharp by comparison to the size of the domain forming at the phase-separation front.  From the definition of the nondimensional speed $U$ in \eqnref{eqn:nondimension_speed} we can see that $U\propto\sqrt{\kappa}$.  Thus a vanishing cost of interfacial free energy ($\kappa\to0$) is equivalent to assuming the front speed is much slower than the spinodal velocity, \textit{i.e.} $U\ll1$.  This assumption of a slow front speed is not especially limiting to our analysis as we are primarily interested in the formation of large stable structures; unlike those found for front speeds approaching the free-front speed $u^*$ which are small and rapidly coarsen \cite{vansaarloos-2003-386}.

The buildup rate of material in a region is
\begin{equation}
	J=\int \partial_t\phi \; d\!x = \int \mathbf{\nabla} \! \cdot \! \mathbf{j} \; d\!x \;,
\end{equation}
where the current $\mathbf{j}$ at a point is determined by the gradient of the chemical potential:
\begin{equation}
	\mathbf{j}(t) = -m \nabla\!\mu(t) \;.
\end{equation}
We're interested in the region of the interface forming directly behind the front at position $x=x_0$.  Since that region is vanishingly small, the rate of material deposited there can be written as a difference of the chemical potential gradients on either side of the front
\begin{eqnarray}
\nonumber J&=&\int (m_M\nabla\!\mu_M - m_S\nabla\!\mu_S)\delta(x-x_0) \; d\!x \\
&=& m_M\nabla\!\mu_M - m_S\nabla\!\mu_S \;.
\end{eqnarray}
Since the mobility ahead of the front is zero, the time dependent rate of deposition for the interface forming at the front is:
\begin{equation}
J(t) = -m_S \nabla\!\mu(t) \;.
\label{eqn:current1}
\end{equation}
The gradient of the chemical potential across the entire forming domain is nearly constant, as can be seen in \figref{timelapse}.  This means that material pulled behind the front is transported all the way across the forming domain and deposited on the second domain.  Any deviation from a linear profile of the chemical potential is of order $\partial_t \phi$, as can be seen from the Cahn-Hilliard equation \eqref{cahn-hilliard} and the fact that $\phi(x,t)$ is nearly stationary inside the forming domain.

The value of the chemical potential at the phase-separation front $\mu_0$ changes with time, but the chemical potential at the other end of the domain is zero.  We can therefore write the chemical potential gradient as
\begin{equation}
  \nabla\!\mu(t) = \frac{\mu_0(t)}{\ell(t)} \;,
	\label{eqn:chemgrad}
\end{equation}
where $\ell$ is the width of the domain forming at the phase-separation front.

The interface at the phase-separation front moves with the front, expanding the domain forming behind it.  This requires that the rate of material deposited into the phase-separation front interface be equal to the difference in concentration across the interface times the front speed
\begin{equation}
	J(t) = u \left[ \phi_0(t) - \phi_{in} \right] \;,
	\label{eqn:current2}
\end{equation}
where $u$ is the (dimensional) front speed, $\phi_0(t)$ is the concentration at the phase-separation front, and $\phi_{in}$ is the mixed-material concentration.

The $M=0$ condition means that none of the wrong-type material can be pushed ahead of the phase-separation front; instead it must be transfered across and deposited behind the forming domain.  This results in the expansion of the second domain by moving the interface at a speed $u_I\leq u$ which trails the phase-separation front.  Taking this into account, and using an argument similar to the one used to derive \eqnref{eqn:current2}, we can write another expression for the rate of material deposition behind the front in terms of material removed and transported to the second interface.  We obtain
\begin{equation}
	-J(t) = (u-u_I(t))2\phi_{eq} \;,
	\label{eqn:current3}
\end{equation}
where $2\phi_{eq}$ is the change in concentration across the fully formed interface, and $\phi_{eq}$ is the equilibrium concentration given by \eqnref{phi_equilibrium}.  For now we'll continue to write $\phi_{eq}$, and eliminate the variable later through nondimensionalization.

The width of the first domain is found by integrating over time the relative speeds of the interfaces at both ends of the domain
\begin{equation}
	\ell(t) = \int_0^t u_I(t) \, dt = u \int_0^t \left( 1 + \frac{\phi_0(t) - \phi_{in}}{2\phi_{eq}} \right) \, dt \;,
	\label{eqn:lambdaint}
\end{equation}
where we eliminated $u_I$ by combining \eqnref{eqn:current2} and \eqnref{eqn:current3}.  We then combine \eqnref{eqn:current1} with \eqnref{eqn:current2}, use the definition of $\nabla\!\mu(t)$ from \eqnref{eqn:chemgrad}, then subtitute the definition of the chemical potential for our model, to find the following alternative expression for the forming domain's width:
\begin{equation}
	\ell(t) = -\frac{m_S}{u}\frac{\mu_0(t)}{\phi_0(t) - \phi_{in}} = -\frac{m_S}{u}\frac{a_S\phi_0(t) + b_S \phi_0^3(t)}{\phi_0(t) - \phi_{in}} \;.
	\label{eqn:lambda}
\end{equation}
Equating \eqnref{eqn:lambdaint} with \eqnref{eqn:lambda} we find the integral equation
\begin{equation}
	\int_0^t \left( 1 + \frac{\phi_0(t) - \phi_{in}}{2\phi_{eq}} \right) \, dt
	= -\frac{m_S}{u^2}\frac{a_S\phi_0(t) + b_S \phi_0^3(t)}{\phi_0(t) - \phi_{in}} \;,
	\label{eqn:intphi}
\end{equation}
for the concentration at the phase-separation front as a function of time.  Differentiation of \eqnref{eqn:intphi} with respect to time yields
\begin{equation}
	1 + \frac{\phi_0 - \phi_{in}}{2\phi_{eq}}
	= -\frac{m_S}{u^2} \left( \frac{b_S\phi_0^2 \left(2\phi_0-3\phi_{in}\right) - a_S\phi_{in}}{\left(\phi_0-\phi_{in}\right)^2} \right) \partial_t \phi_0 \;.
\end{equation}
Therefore
\begin{equation}
	\partial_t \phi_0 = \frac{-u^2 (\phi_0 - \phi_{in})^2 (2\phi_{eq} + \phi_0 - \phi_{in})} { 2 m_S \phi_{eq} \left( b_S \phi_0^2 \left( 2\phi_0 - 3\phi_{in} \right) - a_S \phi_{in} \right)} \;,
	\label{eqn:diffeq}
\end{equation}
where the time dependence of $\phi_0$ is implied.  We nondimensionalize this using Eqs.~(\ref{eqn:nondimension_length}-\ref{eqn:nondimension_phi}), and obtain
\begin{equation}
	\partial_T \Phi_0 =
	\pi^2 U^2 \frac{ (\Phi_0 - \Phi_{in})^2 (2 + \Phi_0 - \Phi_{in}) (3\Phi_{in}^2 - 1)}{2\Phi_0^3 - 3\Phi_0^2\Phi_{in} - \Phi_{in}} \;,
	\label{eqn:diffeqnondim}
\end{equation}
which, for general $U$ and $\Phi_{in}$, can be solved numerically.

For the $\Phi_{in}=0$ case of symmetric inflow concentration,
\eqnref{eqn:diffeqnondim} simplifies to
\begin{equation}
	\partial_T \Phi_0 = - \pi^2 U^2 \left( \frac{1}{\Phi_0} + \frac{1}{2}\right) \;.
	\label{eqn:diffeqsimple}
\end{equation}
The initial concentration must be at equilibrium ($\Phi_0(0)=1$), allowing us to find the analytical solution
\begin{equation}
	\Phi_0(T) = 2 + 2W\left( -\frac{1}{2} \exp\left( \frac{1}{4}\pi^2 U^2 T - \frac{1}{2} \right) \right) \;,
	\label{eqn:diffeqsolution}
\end{equation}
where $W$ is the principle branch of the Lambert~W function.  The Lambert~W function is the solution to the equation $x=W(x)\exp(W(x))$ for complex $x$ and has infinitely many branches.  The principle real branch used for this solution is plotted in \figref{fig:lambertW}, and more information on the Lambert~W function can be found in \cite{corless-1996}.

As explained earlier, our simulations are able to track and record the concentration at the front by inverting the chemical potential.  In \figref{fig:diffeq-phi-of-t} we show that the analytic solution is in excellent agreement with our simulation results for slow phase-separation front speeds.   For faster fronts there is a deviation which is consistent with our assumption of small $U$.

\subsection{Switching Condition}
\begin{center}
	\begin{figure}
		\includegraphics[clip=true, width=\linewidth]{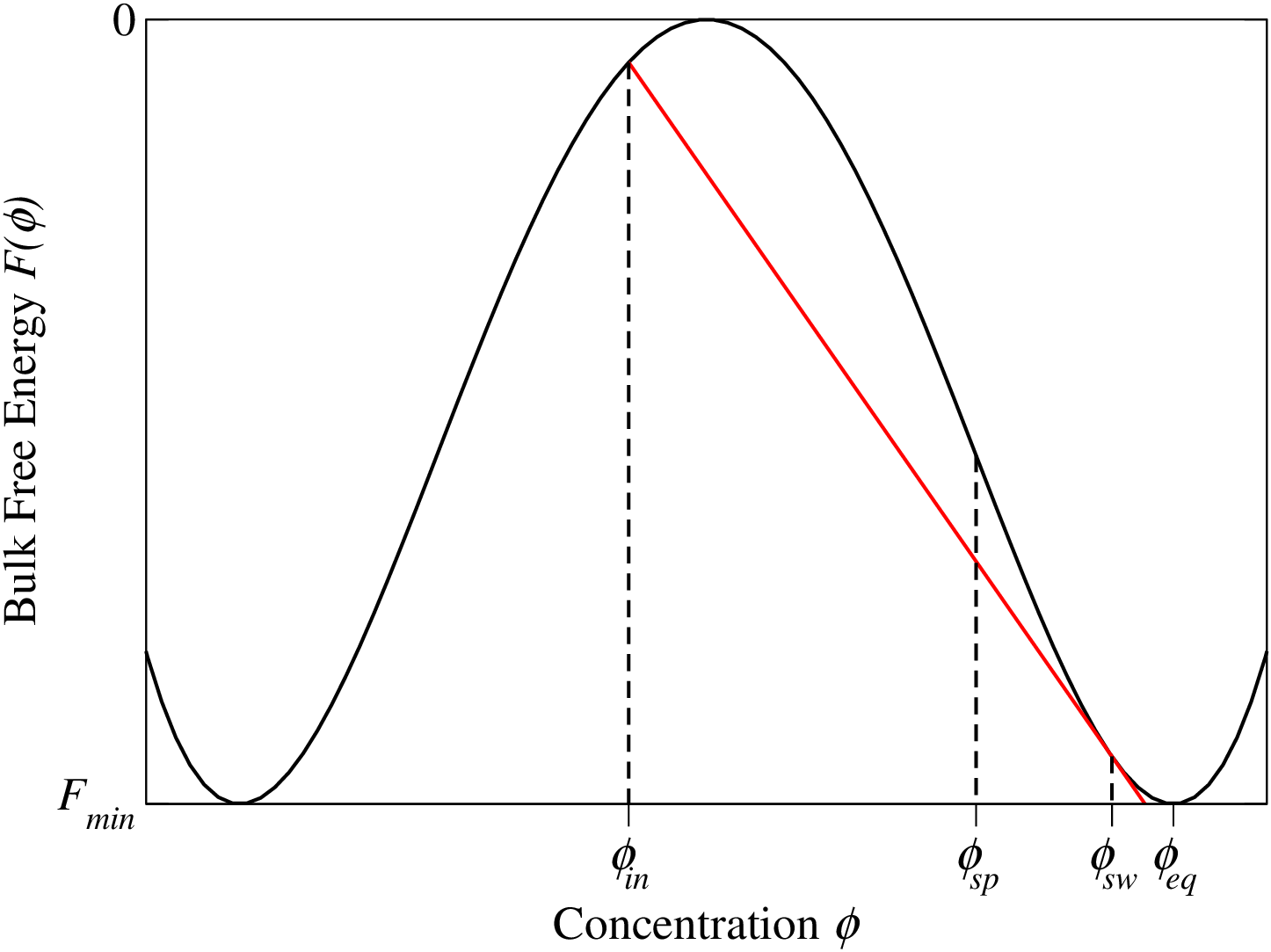}
		\caption{(color online)  Free energy tangent construction for the switching condition for an enslaved phase-separation front.  The nucleation kernel has concentration $\phi_{in}$ and domain type switching will occur when the concentration just behind the front reaches $\phi_{sw}$.  Also marked are the equilibrium ($\phi_{eq}$) and spinodal ($\phi_{sp}$) concentrations.}
		\label{fig:tangent}
	\end{figure}
\end{center}
\begin{center}
	\begin{figure}
		\includegraphics[clip=true, width=\linewidth]{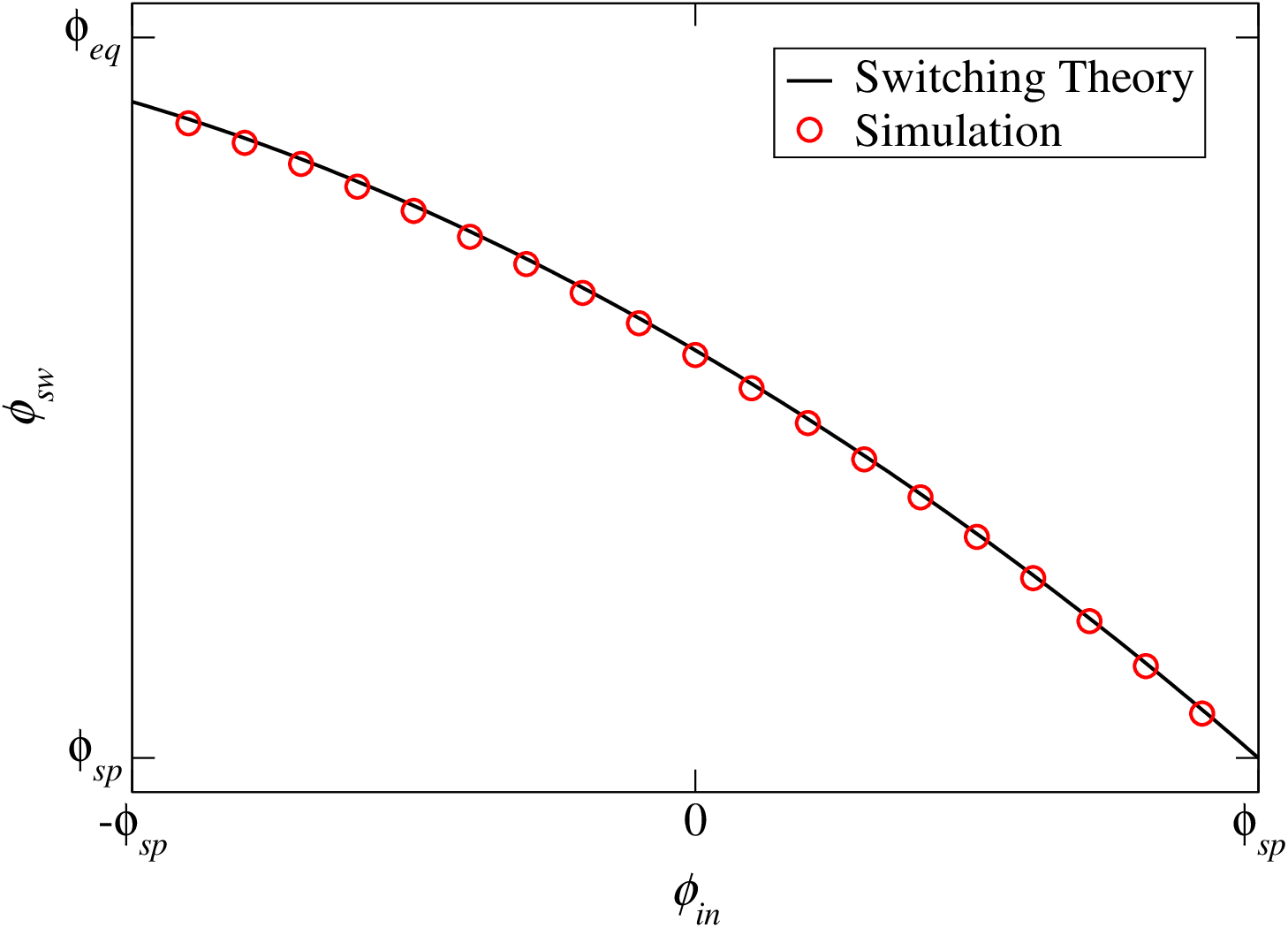}
		\caption{(color online)  Verification of the switching condition for various mixed material concentrations.  The solid line shows the analytic solution for the value of $\phi$ behind the front at which switching occurs.  The circles are switching concentrations recorded from LBM simulations.  The simulation parameters were $M=0$, and $U=6.366\times10^{-3}$.}
		\label{fig:switching}
	\end{figure}
\end{center}
We will now need to determine when a new domain will be formed.  At first glance one might expect that the system will switch when the concentration reaches its spinodal value.  As we saw in \figref{fig:diffeq-phi-of-t}, however, the switch typically occurs much earlier.  Consider an $\mathcal{A}$-rich domain forming behind the front.  Overtaken material can either extend the existing domain by transporting away excess $\mathcal{B}$-type material, or nucleate a new $\mathcal{B}$-rich domain; the system will choose the option that minimizes the total free energy.  We now observe that there is already an interface between the domain forming behind the front and the material the front is overtaking.  Because the interface required to nucleate a new domain is already in place, we only need to consider the bulk free energy for the switching mechanism.  The inflowing material will nucleate a new domain, even without fluctuations, when such a domain formation continuously minimizes the bulk free energy.

This argument is sketched in \figref{fig:tangent}.  The concentration of the potential nucleus of $\mathcal B$-type material forms one point on the free energy curve.  The other point on the free energy curve is the concentration of phase-separated $\mathcal A$ material just behind the front.  If the line-segment intersects the free energy curve at some other point, the free energy can be lowered by dissolving the potential type-$\mathcal B$ nucleus, thus diluting the domain of type-$\mathcal A$ material.  Nucleation will occur if the line-segment connecting these two points on the free energy curve is below the free energy curve.  Thus the system will always switch before the spinodal concentration is reached.

In our case the mixed material ahead of the control-parameter front has a vanishingly small mobility ($M=0$), so the effective concentration of material ahead of the front is simply the concentration of the mixed material.  With that point firmly fixed on the free energy curve, we can predict the critical concentration at which domain switching will occur.

We denote the switching concentration---the concentration at which domain switching due to nucleation occurs---as $\phi_{sw}$.  We just need to find the line which lies tangent to $\phi_{sw}$ on the free energy curve and crosses the nucleation kernel concentration $\phi_{in}$.  The slope of this line is the slope of the free energy curve evaluated at the switching concentration.  This is also the definition of the chemical potential at the switching concentration.
\begin{equation}
	\frac{F(\phi_{sw}) - F(\phi_{in})}{\phi_{sw} - \phi_{in}} = \left. \frac{dF}{d\phi} \right|_{F(\phi_{sw})}
	\!\!\!\!\!\!\!\!\!\!\!\!\!\! = \mu(\phi_{sw}) \;.
\end{equation}
This construction is illustrated in \figref{fig:tangent}, and results in the closed-form solution for the switching concentration:
\begin{equation}
	\phi_{sw} = -\frac{\phi_{in}}{3}\pm\sqrt{\frac{2}{3} - 2\left(\frac{\phi_{in}}{3}\right)^2} \;.
	\label{eqn:switchingcondition}
\end{equation}

We verified \eqnref{eqn:switchingcondition} using our LBM simulations by tracking the peak of the chemical potential near the front during a complete half-cycle and recording the maximum value of the peak for different inflow concentrations $\phi_{in}$.  This is then inverted to find the concentration as a function of the chemical potential.  The concentration at the maximum value of the chemical potential peak is the switching concentration.  The results of this test are shown in \figref{fig:switching}, and demonstrate excellent agreement with the theoretical prediction in the range of $-\phi_{sp}<\phi_{in}<\phi_{sp}$.  Outside this range the system is meta-stable and we do not find alternating domains.

\subsection{Domain Size}
\begin{center}
	\begin{figure}
		\includegraphics[clip=true, width=\linewidth]{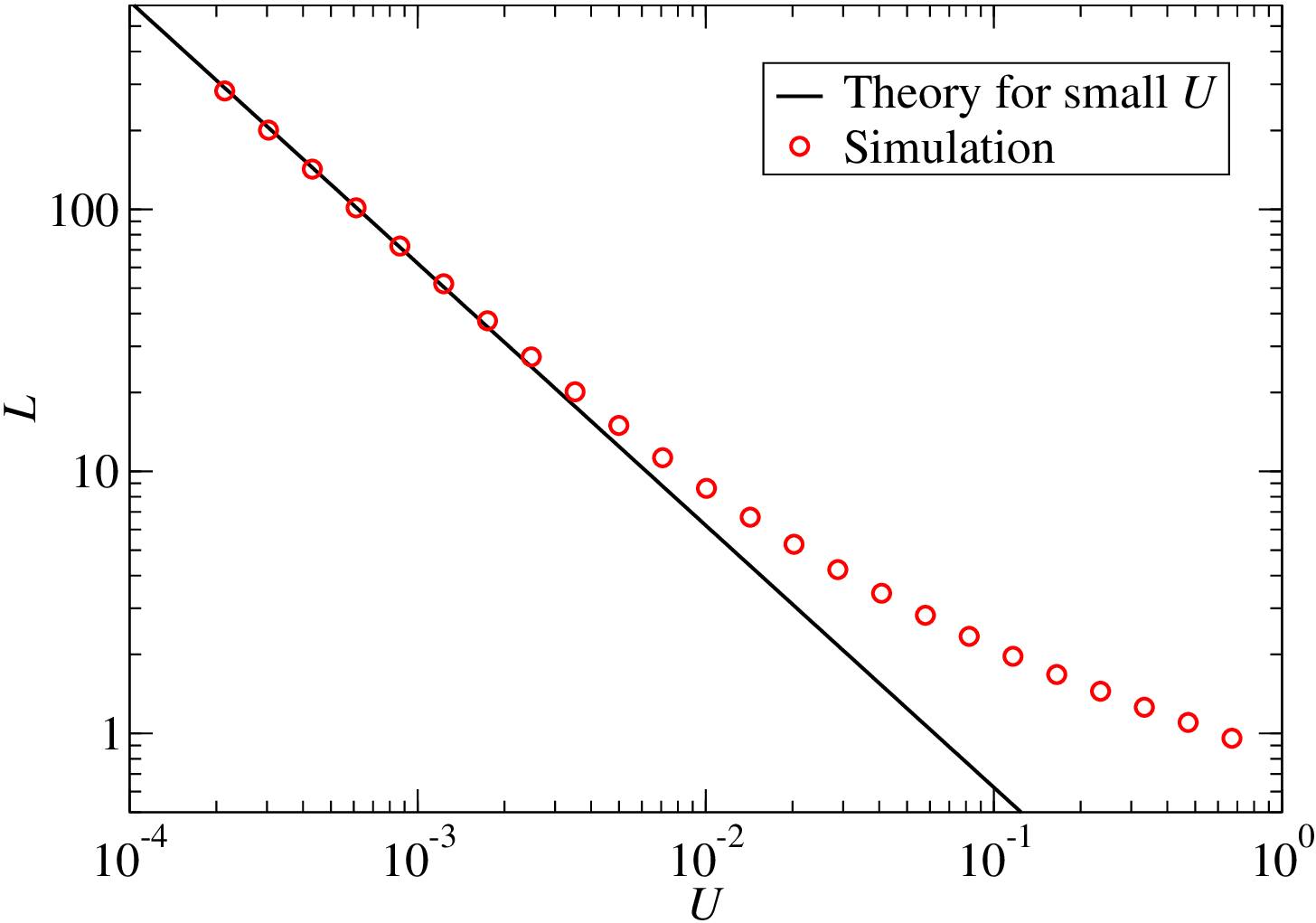}
		\caption{(color online)  Nondimensional domain wavelengths for $M=0$ from lattice Boltzmann method simulations agree very well with the analytical prediction for low phase-separation nondimensional front speed $U$.  Note that the theory contains no adjustable parameters.}
		\label{fig:ul-nobackflow}
	\end{figure}
\end{center}
The expression for the switching concentration in \eqnref{eqn:switchingcondition} is trivially nondimensionalized
\begin{equation}
	\Phi_{sw} = -\frac{\Phi_{in}}{3}\pm\sqrt{\frac{2}{3} - 2\left(\frac{\Phi_{in}}{3}\right)^2}\;,
	\label{eqn:nondimswitchingcondition}
\end{equation}
as is the expected domain wavelength
\begin{equation}
	L = 2 U T_{sw} \;.
	\label{eqn:domainlengthnondim}
\end{equation}
Here $T_{sw}$ is the nondimensional switching time defined as the time when the concentration at the front reaches the switching concentration.  Plugging the switching concentration of \eqnref{eqn:nondimswitchingcondition} for symmetric mixed material inflow ($\Phi_{in}=0$) into the solution of the differential equation for the concentration at the front, presented in \eqnref{eqn:diffeqsolution}, yields
\begin{equation}
	\Phi_{sw} = \sqrt{\frac{2}{3}} = 2 + 2W\left( -\frac{1}{2} \exp\left( \frac{1}{4}\pi^2 U^2 T_{sw} - \frac{1}{2} \right) \right)\;,
\end{equation}
which can be inverted to solve for the switching time:
\begin{equation}
	T_{sw} = \frac{2\left( \sqrt{6} + 6\ln\left( 2-\sqrt{2/3} \right) -3 \right)}{3 \pi^2 U^2} \;.
\end{equation}
With this value for the switching time we can, at long last, explicitly solve for the domain wavelength
\begin{equation}
	L(U) = \frac{4\left( \sqrt{6} + 6\ln\left( 2-\sqrt{2/3} \right) -3 \right)}{3 \pi^2 U} \;,
	\label{eqn:solution}
\end{equation}
as a function of phase-separation front speed for symmetric inflow ($\phi_{in}=0$).  This solution contains no free parameters and only relies on the front speed being much smaller than the spinodal speed.

We verify \eqnref{eqn:solution} by comparing it to data taken from the LBM simulations.  The results are shown in \figref{fig:ul-nobackflow}.  We see that the simulation results agree excellently with our theoretical parameter-free prediction for nondimensional phase-separation front speeds slower than $U\sim10^{-3}$.

The domain wavelengths for asymmetric inflow ($\Phi_{in}\ne0$) can similarly be obtained by numerically solving \eqnref{eqn:diffeqnondim} for the switching time as a function of $\phi_{in}$.

\section{Some Properties of More General Cases}
\begin{center}
	\begin{figure}
		\includegraphics[clip=true, width=\linewidth]{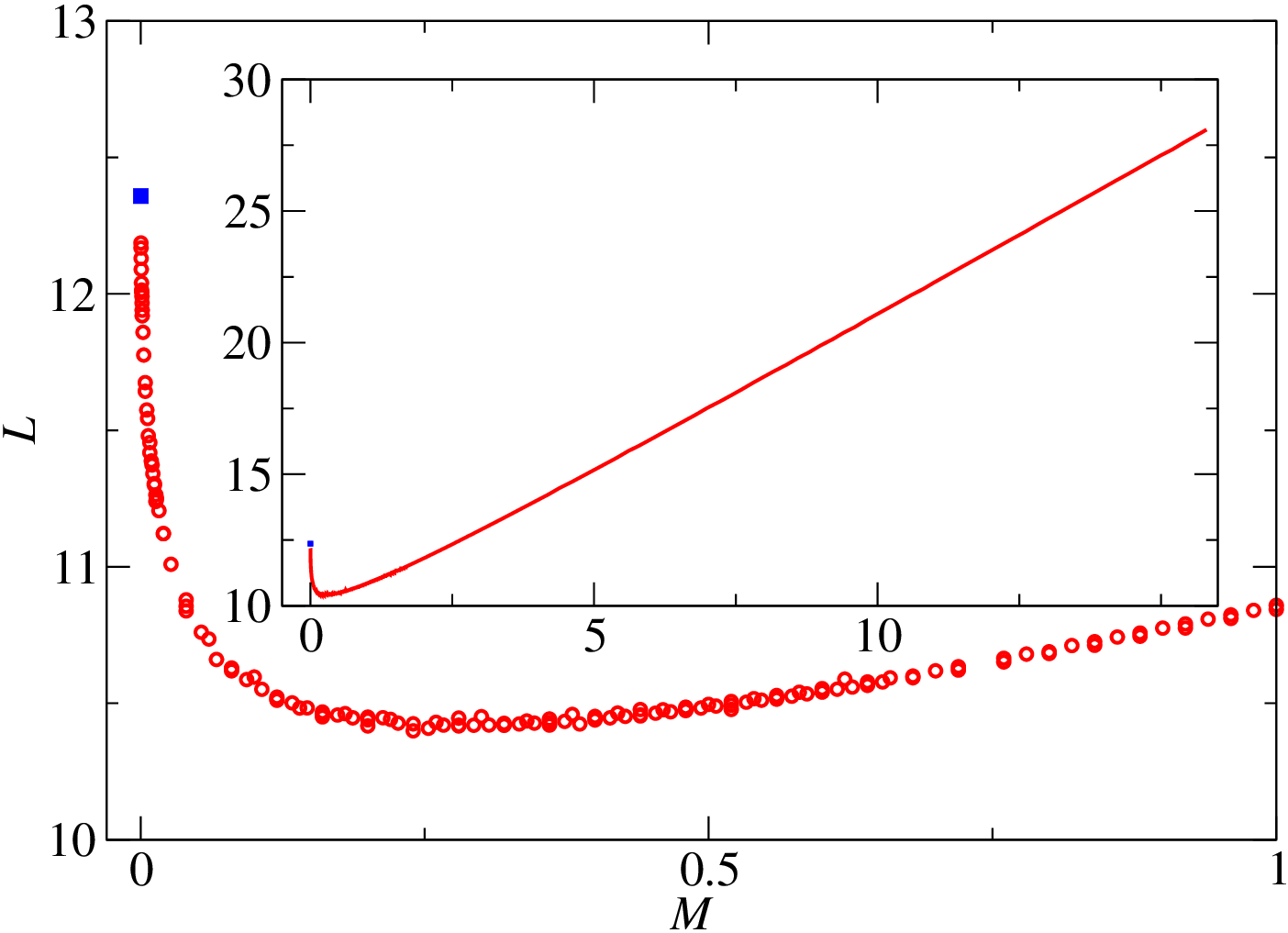}
		\caption{(color online)  Dimensionless domain wavelength $L$ as a function of mobility ratio $M$ as measured from LBM simulations.  The inset plot shows the same data over larger mobility ratio range.  This shows how the family of curves found in the $L$ vs. $U$ plane depends on $M$.  For the red (online) circles, the phase-separation front speed was kept a constant at $U=6.366\times10^{-3}$, while $\kappa$, $m_S$ and $m_M$ were varied.  The blue (online) square is from a simulation of the special case $M=0$ at the same nondimensional speed.  Note this curve is not monotonic.}
		\label{fig:lversusm}
	\end{figure}
\end{center}
We have successfully developed an analytical expression for the morphology formed by an enslaved phase-separation front in the special case where there is no mixed material dynamics and the mixed material is symmetrically composed.  We accomplished this by constructing a differential equation from the analysis of currents which drive the dynamics of phase-separation at the front.  We could only solve this differential equation analytically after the simplification allowed by assuming symmetric mixed material, however the more general case can be tackled numerically.  Along these same lines, one should be able to construct an even more general differential equation which includes dynamics ahead of the control-parameter front.  However, even without such an in-depth analysis we can, at least qualitatively, understand the effects of mixed material mobility on domain formation.

We ``turn-on'' the dynamics ahead of the control parameter front by setting the nondimensional mobility $M$ to some positive value.  This results in a family of curves in the $L$ vs. $U$ plane, each of which are similar in shape to \figref{univscaling}.  To see how this affects the size of domains formed we can choose some slow front speed and perform a series of simulations with varying $M$.  The results of such a series of simulations is shown in \figref{fig:lversusm}.  We observe a rapid but continuous reduction in domain wavelength as the nondimensional mobility is increased from zero, until it reaches a minimum.  We should mention that data of two different kinds of simulation are shown in the graph.  The simulation for $M=0$ where the front is at the inflow boundary, and $M>0$ where the front is firmly inside the lattice.  There is excellent agreement between the methods.

After the sharp reduction levels off, $L$ increases with $M$ and eventually results in the formation of very large domains.  This nonmonotonic behavior is discussed below, followed by an explanation of the effects of the final free nondimensional parameter $A$.

\subsection{Small Mixed Mobility (M${\mb\ll}$1)}
If the dynamics of material ahead of the front are very slow by comparison to the separated region, the currents are almost identical to the case of no mixed-material dynamics.  Such a case could be physically realized for a system where the mobility strongly increases with temperature and we have a lower critical point.  In this case, very little material is allowed to build up ahead of the front.  Currents ahead of the front are then vanishingly small.  What does change, due to this small buildup ahead of the front, is that the front is now overtaking a slightly larger nucleus of material.  This slightly larger nucleus will make domain type switching more energetically favorable.

We can see the effect a small increase in nucleus volume has on the switching condition by re-examination of \figref{fig:tangent}.  A slightly larger nucleation kernel has the effect of shifting $\phi_{in}$ towards the opposite material type.  This results in a shallower tangent line construction which contacts the free energy curve closer to the equilibrium concentration.  This means that the concentration of material just behind the front cannot be diluted as much before domain type switching occurs, and earlier domain type switching results in smaller domains.

\subsection{Large Mixed Mobility (M${\mb\gg}$1)}
\begin{center}
	\begin{figure}
		\includegraphics[clip=true, width=\linewidth]{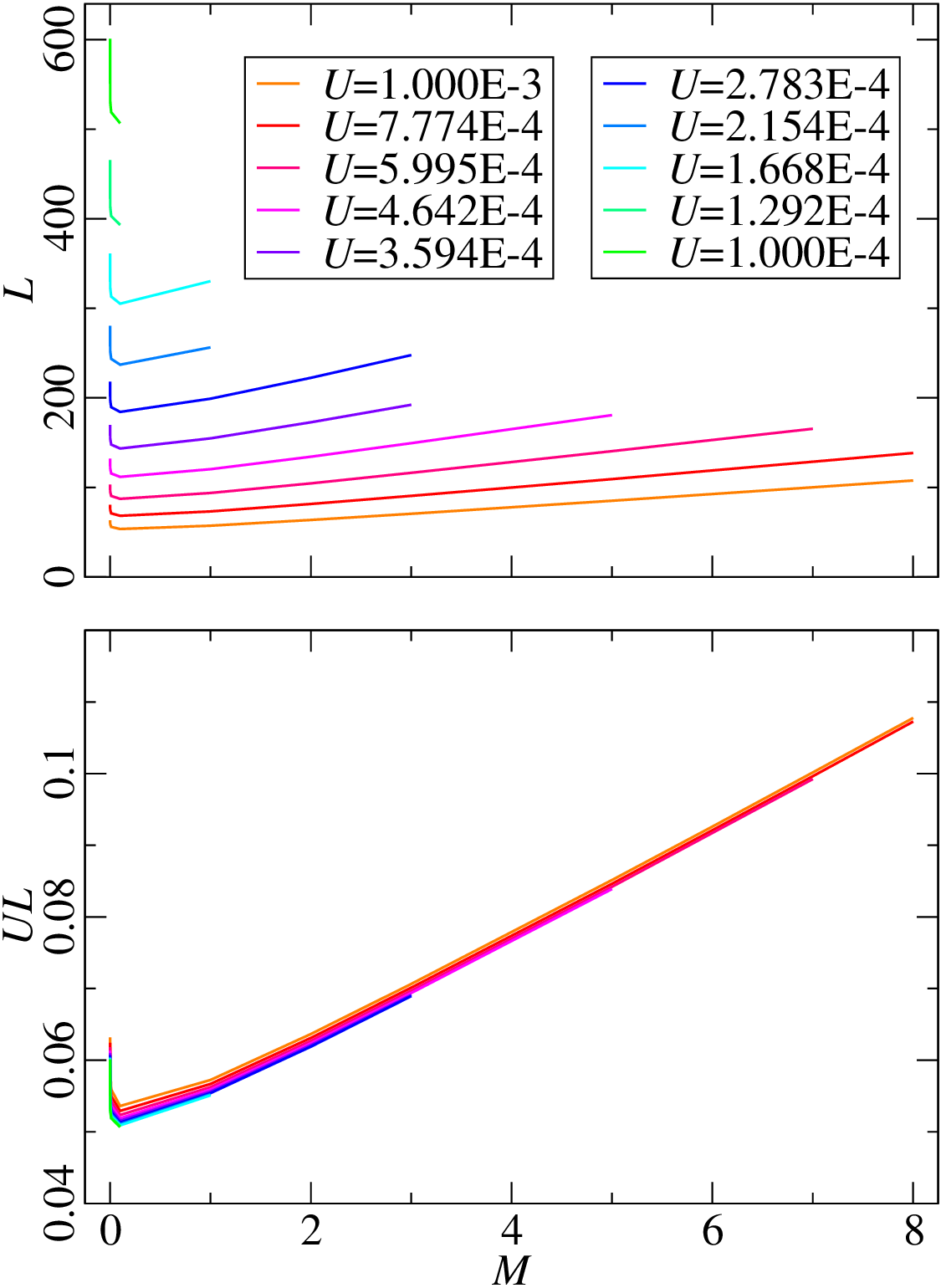}
		\caption{(color online)  Domain length scaling is universal for changes in front speed and mobility ratios in the slow front speed limit.  The upper graph shows that the nondimensional domain wavelengths $L$ and front speed $U$ both vary by an order of magnitude, yet retain similar curve shapes.}
		\label{fig:univscaling2}
	\end{figure}
\end{center}
This is a case that is easily realizable in real systems.  For strongly phase-separated regions the mobility can become very small compared to the mixed material.  As we saw from the results of the simulations shown in \figref{fig:lversusm}, for large $M$ the domain wavelengths become very large.  This happens because diffusion is rapid in the mixed material region, and large amounts of material can build up ahead of the front.  When the domain at the front switches type, this buildup floods into the new domain, thereby making it very large.  In the limit of large $M$, the domain is predominantly formed from this material.  Since this is effectively the same as halting the dynamics behind the front, we can understand this further by examining just the mixed material dynamics ahead of the front.

In the reference frame of the front, material pushed ahead of the front is governed by the drift-diffusion equation
\begin{equation}
	\partial_{t} \phi + \nabla(u\phi) = \nabla m \nabla \phi \;,
	\label{eqn:drift-diffusion}
\end{equation}
which is defined for $x>x_0$.  Here we have neglected the interfacial term and absorb the $a$ parameter into the mobility.  The boundary condition at the front at $x=x_0$ is
\begin{equation}
	J(t) = u [ \phi(x_0,t) - \phi_{eq} ] \;,
\end{equation}
which is the current of material rejected by the forming domain, analogous to the result of \eqnref{eqn:current2}.  This boundary condition results in a buildup of material which will eventually lead to switching.  These two equations are sufficient to define the material dynamics ahead of the front.  We now nondimensionalize these equations to recover their parameterless versions.

We introduce the general length and time scales
\begin{equation}
	x={\mathcal X}\hat{x}\;\mbox{, and}\;t={\mathcal T}\hat{t}\;.
\end{equation}
It should be noted that ${\mathcal X}$ and ${\mathcal T}$ are not the same as the spinodal wavelength and time we used in the earlier nondimensionalization.  The drift diffusion equation then becomes
\begin{equation}
	\frac{1}{{\mathcal T}}\partial_{\hat{t}} \phi
		+ \frac{1}{{\mathcal X}} \widehat{\nabla}
			\left(
				\frac{{\mathcal X}}{{\mathcal T}} \hat{u} \phi
			\right)
		= \frac{1}{{\mathcal X}} \widehat{\nabla}
			\left(
				m \frac{1}{{\mathcal X}} \widehat{\nabla} \phi
			\right)
		\;,
\end{equation}
which we rewrite as
\begin{equation}
	\partial_{\hat{t}} \phi + \widehat{\nabla} (\hat{u} \phi) = \widehat{\nabla} \left( \frac{m{\mathcal T}}{{\mathcal X}^2} \widehat{\nabla} \phi \right)\;.
\end{equation}
For the boundary condition
\begin{equation}
	J(t) = \frac{{\mathcal X}}{{\mathcal T}} \hat{u}(\phi(x_0,t)-\phi_{eq})\;,
\end{equation}
we obtain the nondimensionalized boundary current:
\begin{equation}
	\hat{J}(\hat{t}) = \hat{u}(\phi(x_0,t)-\phi_{eq})\;.
\end{equation}
We now choose the length and time scales such that
\begin{equation}
	1 \equiv \hat{u} = \frac{\mathcal T}{\mathcal X}u \;\mbox{, and}\; 1 \equiv m\frac{\mathcal T}{\mathcal X^2}\;.
	\label{nondim_scaling}
\end{equation}
This results in the parameterless equation of motion 
\begin{equation}
	\partial_{\hat{t}} \phi + \widehat{\nabla}\phi = \widehat{\nabla}^2 \phi\;,
\end{equation}
and boundary condition
\begin{equation}
	\hat{J} = (\phi(x_0,t)-\phi_{eq})\;.
\end{equation}

While we do not have the analytical solution of this differential equation, we know that the solution exists and will result in a nondimensional switching time $\hat{t}_{sw}$.  We use this to replace the dimensional switching time appearing in \eqnref{eqn:domainlength}, which gives the domain wavelength
\begin{equation}
	\lambda = 2 u {\mathcal T} \hat{t}_{sw}\;.
\end{equation}
We now evaluate ${\mathcal T}$ in terms of $m$ and $u$ from \eqnref{nondim_scaling}:
\begin{equation}
	{\mathcal T} = \frac{m}{u^2} \;.
\end{equation}
This uncovers the expected domain wavelength
\begin{equation}
	\lambda = 2 \frac{m}{u} \hat{t}_{sw} \propto \frac{m}{u}\;.
\end{equation}
We can re-express this result in terms of the earlier nondimensionalization of Eqs.~(\ref{eqn:nondimension_length}-\ref{eqn:nondimension_phi}) as
\begin{equation}
	L \propto \frac{M}{U} \;.
\end{equation}
Thus we uncover the expected scaling behavior of the domain wavelength for the simultaneous limit of small $U$ and large $M$.

Interestingly, the scaling behavior of 1-D coarsening of a homogeneous quench is logarithmic in time, therefore enslaved phase-separation fronts can build large domains much more effectively than the phase-ordering which follows spinodal decomposition.  In principle, the speed of the front can be controlled, and thus the production of highly ordered, depth-dependant structures becomes possible.

We have already seen from the analytical result for $M=0$ in the limit as $U \to 0$ that the domain wavelength is proportional to the inverse of the front speed ($L\propto U^{-1}$) which matches the previous scaling argument for the opposite case of $M \to \infty$.  The scaling argument reveals that $LU$ vs $M$ will result in universal scaling in the limit of very slow fronts.  We test this by performing simulations on a selected range of sufficiently slow front speeds $U\leq10^{-3}$, and a range of mobilities.  The results of these simulations are shown in \figref{fig:univscaling2} revealing the large difference in domain wavelengths in the $U$ vs $M$ plot which then collapse reasonably well in the $LU$ vs $M$ plot.

To improve the collapse, even slower simulations would be required, however this becomes too computationally expensive.  For the measurements presented here we required simulations of more than $10^5$ lattice sites simulated for more than $10^8$ iterations.  Such simulations take about one week on a Xeon powered Linux workstation.

\subsection{$A$ and $M$ at the Front Boundary}
\begin{center}
	\begin{figure}
		\includegraphics[clip=true, width=\linewidth]{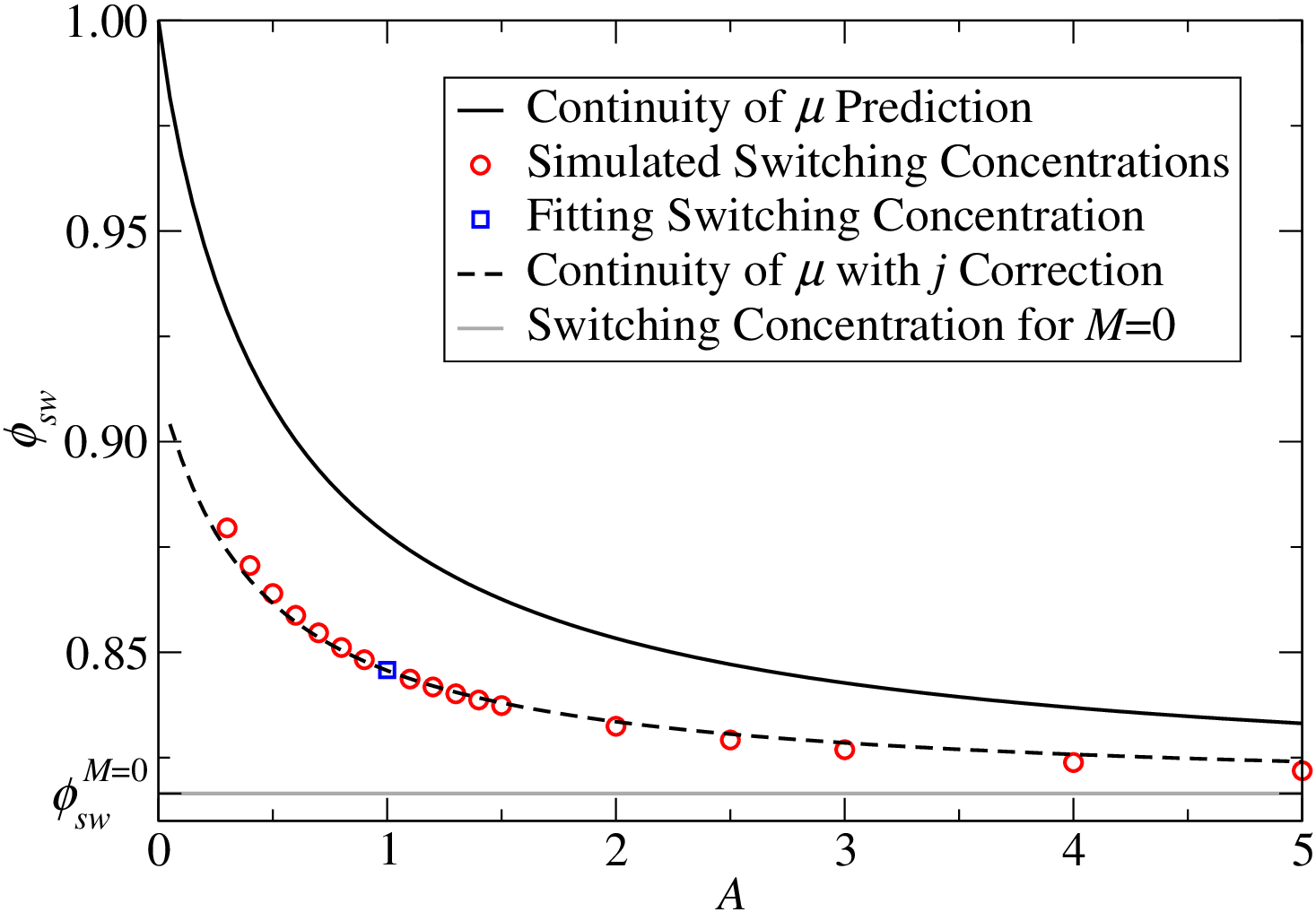}
		\caption{(color online)  Switching concentration for large nondimensional mobility $M$ as a function of control parameter height $A$.  The predicted curves are calculated by assuming perfect continuity of the chemical potential across the front, and using this to predict the size of the nucleation kernel in the switching condition.  The dashed line includes the addition of a current $j$ which is caused by the dynamics ahead of the front, and is zero in the $M=0$ case.  The simulation used a nondimensional mobility ratio $M=5$, and front speed $U=10^{-4}$.}
		\label{fig:A-switching}
	\end{figure}
\end{center}
The relative quench depth $A$ is the last relevant parameter which needs to be considered in determining morphology formed by slow enslaved phase-separation fronts in 1D.  The control parameter in either a purely mixed ($A_M$) or purely separating ($A_S$) system is usually not a function of position and can be absorbed into the mobility.  In fact, if we were to consider the mixing region ahead of the front as being entirely detached from the separating region behind the front, we would be able to completely nondimensionalize away all free parameters, resulting in separate scale-invariant equations of motion.  However, due to the shared boundary at the front linking the two regions, the change of the control parameter $A$ across the front cannot be ignored.  This means that changing the control parameter ratio $A$ has different consequences for the resulting morphology than does changing the mobility ratio $M$.

The first consequence of changing the order parameter $A$ has for the dynamics of phase-separation is for the material which gets pushed ahead of the control parameter front.  Recall that the chemical potential is continuous across the front ($\mu_S=\mu_M$).  Ignoring for the moment the interfacial contribution to the chemical potential, we can write the continuity of the the chemical potential as 
\begin{eqnarray}
	\nonumber \phi_S^3 - \phi_S &=& A\phi_M \;, \\
	\Rightarrow \phi_M &=& \frac{1}{A}\left(\phi_S^3 - \phi_S\right) \;.
	\label{mu_continuity}
\end{eqnarray}
This means that the concentration just ahead of the front and the concentration just behind the front are tied together.  This is important because the switching condition, even for very large $M$, is partly dependent on the concentration of material just ahead of the front.  For $A\ll1$, the concentration ahead of the front becomes larger, causing the switching concentration to approach the equilibrium concentration, resulting in very early switching.  On the other hand, for $A\gg1$ the concentration ahead of the front becomes close to the symmetrically mixed concentration, resulting in switching taking longer to occur.

If we simply use $\phi_M$ in \eqnref{mu_continuity} as an effective $\phi_{in}$ in the earlier derived switching condition of \eqnref{eqn:switchingcondition}, we discover a new switching condition for large $M$ which is dependent only on $A$.  This predicted switching concentration is plotted as the solid line in \figref{fig:A-switching}, along with switching concentrations measured from our LBM simulations.  Since the switching condition assumed no dynamics ahead of the front it is not surprising that this prediction is inadequate for the opposite scenerio.  If we take into account the additional current $j$ across the front due to the dynamics ahead of the front, and account for this in the effective $\phi_{in}$ in the switching condition
\begin{equation}
	\phi_{in}^{eff} = \frac{1}{A}\left(\phi_S^3 - \phi_S - j \right) \;,
\end{equation}
we can measure this current at one value of $A$ and use it to fit the rest of the curve.  This new prediction is the dashed line in \figref{fig:A-switching}, with the $A=1$ switching concentration used as the fitting value.  This new prediction agrees well, although not perfectly, with the simulated values, showing that the current across the front at switching seems to scale as the inverse of $A$.

\section{Outlook}
Phase-separation front driven morphology is a rich and complex subject which has significant potential for new research and applications.  In this paper we focused on the simplest case: a one-dimensional binary mixture.  For abrupt control-parameters we found that a regular alternating morphology is formed.  We have shown that, complex though the subject may be, certain important aspects of enslaved phase-separation fronts can be understood well enough to theoretically predict the morphology.

Using our model of a binary mixture with an abrupt control parameter front, we were able to determine the effect of all relevant free parameters on the size of domains formed in a 1D system.  We verified our predictions by comparing them to LBM simulations of our model.

We were able to reduce the number of relevant parameters from twelve to four.  We achieved this by rescaling length and time with the spinodal wavelength and time and the concentration by the equilibrium value of the order parameter, and by making some simple physical assumptions.

Analytical solutions are few and far between for the theory of morphology formation.  Choosing the nondimensional front speed as the primary parameter of interest, we consider the radical simplification of a front moving into material which has no mixed-material dynamics.  This fixes the other free parameters and allows us to analytically determine the dynamics of phase-separation induced by a slow front.

We first find the time-dependence of the concentration at the front.  We then determine the mechanism for domain switching.  While typical nucleation arguments are heavily dependent on the interfacial free energy our switching condition was found to depend purely on the bulk free energy.

This gives us an analytical prediction of the formed which is in remarkable agreement with the numerical results.  Additionally, this solution functioned as a stepping stone for understanding the effects of the remaining parameters.

By allowing dynamics in the material ahead of the front we discovered a nonmonotonic dependence of the domain wavelength on the nondimensional mobility.  Allowing a small amount of material to be pushed ahead of the front actually decreased domain wavelength, whereas allowing a large buildup of material ahead of the front resulted in huge increases in domain wavelength.  The former is understood by an extension of the switching condition, and the later is explained by a simple scaling argument; when the mobility in the mixing region is much greater than separating region, the domain wavelength formed by the front should scale with the same dependence as the length scale in the mixing region.  This argument resulted in the discovery that, for very slow front speeds, there is a single universal scaling curve of the nondimensional domain wavelength as a function of the nondimensional mobility and front speed.

Domain type switching is very well understood for the case of no mixed material dynamics, and we've explained the effect of very small mixed material mobility.  The effect of large mixed material mobility on the switching condition is uncovered in the analysis of the quench depth $A$ of the enslaved front.  Modification of the switching condition is accomplished by including the continuity of the chemical potential, and a current $j$ which is inversely proportional to $A$.

The effect on one-dimensional morphology formation of three of the relevant nondimensional parameters is now well understood.  We now briefly mention several research avenues that lead on from the results presented in this paper.

We are currently examining higher-dimensional enslaved control parameter fronts.  Two-dimensional fronts open the possibility for the formation of stripe morphologies which are perpendicular or at an angle to the control parameter front.  Additionally we find other exotic morphologies like polka-dot lattices and ovoid domains.  Also, dimensionalities of two and higher allow hydrodynamics to play a roll in the phase-separation dynamics.  In three-dimensions, enslaved fronts can induce a rich family of morphologies depending on the properties of the material and the front.  Perhaps the most intriguing possibility is to be able to control the material and front well enough to reliably switch morphologies during front traversal, resulting in highly-ordered, highly-inhomogeneous composite materials.

What also remains is to examine more general enslaved fronts.  The most straightforward extension is to attempt to develop an analytical solution to the abrupt one-dimensional enslaved front for all three free parameters $U$, $A$, and $M$.  A careful accounting of the additional currents cased by dynamics ahead of the front should allow an extension of the differential equation describing the dynamics at the front, presented in \eqnref{eqn:diffeqnondim}, for this more general abrupt one-dimensional enslaved front.

Then we will relax the condition of an abrupt control parameter front to obtain an extended front.  We expect that there will be little change while the width of an extended control parameter front is on the order of the interface width, but extended fronts which approach the size of the forming domain may result in new and interesting phase-separation dynamics, even in 1D.

\end{document}